\documentclass[10pt,twocolumn,twoside]{IEEEtran}

\usepackage{latexsym,psfig,times,epsfig,epsf,amsmath,amssymb,amsfonts,graphicx, multirow, subfigure, psfrag, simplemargins,color,url,epstopdf,algorithm,algorithmic,setspace}

\newcommand{\tabincell}[2]{\begin{tabular}{@{}#1@{}}#2\end{tabular}}

\newtheorem{lemma}{\textbf{Lemma}}

\newtheorem{theorem}{\textbf{Theorem}}

\newtheorem{definition}{\textbf{Definition}}

\newtheorem{assumption}{\textbf{Assumption}}
\newtheorem{remark}{\textbf{Remark}}

\def\QED{\mbox{\rule[0pt]{1.5ex}{1.5ex}}}
\def\proof{\noindent\hspace{2em}{{\itshape Proof : }}}
\def\endproof{\hspace*{\fill}~\QED\par\endtrivlist\unskip}
\def\@begintheorem#1#2{\tmpitemindent\itemindent\topsep 0pt\rmfamily\trivlist
       \item[\hskip \labelsep{\indent\itshape #1\ #2:}]\itemindent\tmpitemindent}
\def\@opargbegintheorem#1#2#3{\tmpitemindent\itemindent\topsep 0pt\rmfamily \trivlist
       \item[\hskip\labelsep{\indent\itshape #1\ #2\ \rmfamily(#3)}]\itemindent\tmpitemindent}
\def\@endtheorem{\endtrivlist\unskip}

\newcommand{\ls}[1]
   {\dimen0=\fontdimen6\the\font
    \lineskip=#1\dimen0
    \advance\lineskip.5\fontdimen5\the\font
    \advance\lineskip-\dimen0
    \lineskiplimit=.9\lineskip
    \baselineskip=\lineskip
    \advance\baselineskip\dimen0
    \normallineskip\lineskip
    \normallineskiplimit\lineskiplimit
    \normalbaselineskip\baselineskip
    \ignorespaces
   }

\renewcommand{\vec}[1]{\boldsymbol{#1}}

\title{On Maximizing Sensor Network Lifetime by \\Energy Balancing}

\begin{document}
\author{\IEEEauthorblockN{Rong~Du\IEEEauthorrefmark{1},~Lazaros~Gkatzikis\IEEEauthorrefmark{1},~Carlo~Fischione\IEEEauthorrefmark{1},~Ming~Xiao\IEEEauthorrefmark{2}}\thanks{This work is supported by the Wireless@KTH Seed Project LTE-based Water Monitoring Networks and ICT EIT Lab project I3C.}\\
\IEEEauthorblockA{\IEEEauthorrefmark{1} Automatic Control Department, \IEEEauthorrefmark{2} Communication Theory Department\\
KTH Royal Institute of Technology, Stockholm, Sweden }\\
\IEEEauthorblockA{Email: \{rongd, lazarosg, carlofi, mingx\}@kth.se}\\
}

\maketitle

\begin{abstract}
	Many physical systems, such as water/electricity distribution networks, are monitored by battery-powered Wireless Sensor Networks (WSNs). Since battery replacement of sensor nodes is generally difficult, long-term monitoring can be only achieved if the operation of the WSN nodes contributes to a long WSN lifetime. Two prominent techniques to long WSN lifetime are i) optimal sensor activation and ii) efficient data gathering and forwarding based on compressive sensing. These techniques are feasible only if the activated sensor nodes establish a connected communication network (connectivity constraint), and satisfy a compressive sensing decoding constraint (cardinality constraint). These two constraints make the problem of maximizing network lifetime via sensor node activation and compressive sensing NP-hard. To overcome this difficulty, an alternative approach that iteratively solves energy balancing problems is proposed. However, understanding whether maximizing network lifetime and energy balancing problems are aligned objectives is a fundamental open issue. The analysis reveals that the two optimization problems give different solutions, but the difference between the lifetime achieved by the energy balancing approach and the maximum lifetime is small when the initial energy at sensor nodes is significantly larger than the energy consumed for a single transmission. The lifetime achieved by the energy balancing is asymptotically optimal, and that the achievable network lifetime is at least $50$\% of the optimum. Analysis and numerical simulations quantify the efficiency of the proposed energy balancing approach.

\end{abstract}

\section{Introduction} \label{sec:intro}

Wireless sensor networks (WSNs) are being used to monitor critical infrastructures in smart cities, such as water distribution networks, tunnels, bridges, and towers. Since sensor nodes are generally power limited, and battery replacement is difficult or even impossible, network lifetime is an important performance metric \cite{dietrich2009lifetime}. Several approaches have been proposed  to prolong network lifetime and hence to enable long-term monitoring. For example, sensor nodes can form clusters, where participating nodes  take turn to act as cluster-head to balance the energy consumption of the nodes \cite{heinzelman2000energy,hoang2014cluster,leu2015energy}. The nodes can optimize routing \cite{chen2013near,imon2013rasmalai} or use multi-hop short range communication \cite{song2009maximizing} to save energy in the data transmission. Moreover, event-trigger mechanisms \cite{halkes2005comparing} can be used to reduce the transmitted data volume. The sensor nodes can also be put into sleep or idle mode to save energy \cite{Degirmenci2014maximizing,rong2015JSAC}. The methods to be used for energy saving should depend on the characteristics of the monitoring applications.

In this paper, we consider the case of using densely deployed sensors to monitor an area where node replacement is difficult. Such a dense sensor network has the following benefits:

\begin{itemize}
	\item better detection of events;
	\item robustness to sensor failure and measurement errors because of the availability of redundant sensor nodes. Thus, the network operation is ensured even when some sensor nodes fail;
	\item reduced energy consumption in data transmission by exploiting multi-hop short range communication. Thus, network lifetime is increased.
\end{itemize}
Consequently, even though dense networks introduce a higher installation cost, they substantially reduce the maintenance cost in return, and, most importantly, may provide better monitoring performance.

We consider to use data compression in the data gathering process, together with a sleep/awake mechanism for the sensing process, to prolong lifetime for such a dense sensor network. A natural question is whether the usual approach of energy balancing, i.e., preferably use
the nodes with more residual energy~\cite{abdelsalam2012toward}, would be a viable choice for maximizing network lifetime in this context.

As the sensor nodes are densely deployed, their measurements exhibit spatial correlations. Such correlations enable us to use compressive sensing (CS) to accurately estimate the state of the monitored infrastructure with a minimal number of measurements  \cite{candes2006robust,xu2013efficient}. Therefore, one may adopt a CS-based data gathering scheme, such that in every timeslot only a portion of sensor nodes is activated  to sense and transmit data hop-by-hop to the sink nodes. It follows that the expected monitoring performance of such a system can be guaranteed by CS while its energy efficiency  can be improved by turning off the rest of the sensor nodes.

In our previous works \cite{rong2015energy,rong2015JSAC}, we proposed a CS  based sensor activation scheme based on energy balancing for dense WSNs to monitor water distribution networks. The devised data gathering scheme activates only a few connected sensor nodes for sensing and data transmission, to reduce the overall energy consumption and so that the monitoring performance is guaranteed even under sensor failures. However, whether that algorithm (or any energy-balancing based one) can achieve the maximum WSN lifetime is an open issue that has not been investigated before in the CS context. In summary, we address this fundamental issue of whether, in
the considered WSN scenario, the energy balancing problem is equivalent to lifetime maximization. The main results of the paper are as follows:
\begin{itemize}	
	\item In order to provide insight on the complexity and the structure of the Lifetime maximization Problem, we cast it as a Multi-dimensional Knapsack problem , for which a rich literature on solution approaches exists.
	\item We provide an easy to calculate upper bound of maximum lifetime based on a transformation to a maximum flow problem, as shown by Theorem~\ref{thm:lifetimeupperboundflow} in Section~\ref{sec:performance} A.
	\item We propose an algorithm that gives an approximate solution to the maximum lifetime problem. The algorithm is based on the solution to an energy balancing problem. We show that such an algorithm is asymptotically optimal (as given by Theorem~\ref{prop:multiply} in Section~\ref{sec:performance} C) and the worst case approximation ratio (the ratio of the lifetime achieved by the algorithm to the optimal lifetime) is 50\% (as shown in Theorem~\ref{thm:approxratio}  in Section~\ref{sec:performance} C). The asymptotic optimality and the approximation ratio of the algorithm constitute major original contributions of this paper because for maximum lifetime problems with connectivity and cardinality constraints there are no known optimality bounds. 
\end{itemize}

The organization of the rest of the paper is as follows. We provide an overview of related literature in Section~\ref{sec:relatework}. The formulation of the lifetime maximization problem and the energy balancing problem are described in Section~\ref{sec:problem}. In Section~\ref{sec:performance}, the performance of the proposed algorithm in terms of network lifetime is analyzed. 
Numerical evaluations are provided in Section~\ref{sec:sim}. The conclusions of this work are presented in Section~\ref{sec:conclusion}.

\section{Related Work}\label{sec:relatework}

\subsection{Lifetime maximization by flow approximation}


The lifetime of a network greatly depends on the residual energy of the participating nodes. There are different models for the energy consumption of sensor nodes. In \cite{Chang04}, the energy consumption is linearly related to the receiving power, transmitting power and data transmission rate, and the expected lifetime of a sensor node is defined as the ratio of the energy capacity of the node and the average energy consumption. In~\cite{jarry2006optimal}, a slice model is used where the monitored area is partitioned into slices, each of which contains all the nodes that have the same hop distance from the sink node. In this model, the energy consumption of the nodes depends also on the distance (hop count). However, the distances of a node to any node in the same slice are considered to be the same. In~\cite{Degirmenci2014maximizing}, the energy consumptions of the nodes in sleep mode are assumed to be 0, whereas that of the active nodes follows an independent and identical distribution. In~\cite{Cassandras2014}, a non-linear dynamic energy consumption model, called kinetic battery model, is used. In~\cite{shi2016adaptive}, given the fixed topology of a WSN, the energy consumption rate of the working sensor nodes is considered constant, whereas the consumption of the sleeping sensor nodes is 0. Similar to~\cite{shi2016adaptive}, we normalize the energy consumption of the active sensor nodes to 1 and set that of the nodes in sleep mode equal to 0.

One common way to prolong network lifetime is by reducing energy consumptions of each node. For example, in the event-trigger based approaches \cite{halkes2005comparing}, one could reduce the sampling rate of the sensors, to save energy of sensing and data transmission. Notice that for general networks, the data transmission constitutes the major part of a node's energy consumption. Thus, several solutions have been considered to reduce nodal energy consumption due to data transmission, such as controlling the transmission power \cite{lin2006atpc,fan2015delay} and compressing the measurements \cite{incebacak2015optimal, karakus2013analysis} to be transmitted. Besides, harvesting energy from ambient environment \cite{cammarano2013energy,berger2015sustainable,shi2016adaptive}, such as solar energy and vibration, or transmitting energy to the nodes wirelessly \cite{xie2012renewable,rong2016lifetime}, can also extend the lifetime of a node, but are beyond the scope of this work. 

Besides the nodal perspective, prolonging lifetime from the network perspective, e.g., by optimized routing, has been widely studied. In this context, the classic flow  approximation is commonly used \cite{Chang04,jarry2006optimal}. In particular, the energy budget of each node is represented as the number of flows that can pass the node, which is referred to as `vertex capacity'. Then, finding the route in each wireless communication timeslot to maximize network lifetime is equivalent to finding the maximum flow from source node to sink node. In the seminal work \cite{Chang04}, the energy consumption of the network has been modelled as a function of the traffic flow routing decisions. Then the problem is cast as a linear programming problem. In a similar network setting, where every sensor node can either transmit its data to its neighbor with low energy cost, or transmit data directly to the sink node with high energy cost, maximizing network lifetime is equivalent to flow maximization and energy balancing \cite{jarry2006optimal}. In such scenario, energy balancing has been used to maximize network lifetime \cite{zhang2009balancing,ren2011ebrp}. Another way to balance the energy consumption is rotating the working period of sensor nodes, i.e., allowing some sensor nodes to sleep without sacrificing in the monitoring performance. For instance, Misra et. al.~\cite{misra2009rotation} have considered finding different connected dominating sets of the WSN to prolong network lifetime. In each timeslot, only the sensor nodes in the connected dominating set are active and the other nodes are put into sleep. To rotate the working period of the nodes, it is desired to find the maximum connected domatic partition, which divides the WSN into as many as possible disjoint connected dominating sets. A similar problem has been considered in~\cite{shi2016adaptive}, where the sensor nodes have the energy harvesting ability. Thus, the working schedules of the connected dominating sets are also taken into consideration.

Compared to the WSN scenarios mentioned above, CS considered in this paper introduces a cardinality constraint. It is an open question how the problem of lifetime maximization of such WSN is related to the energy balancing problem.
Another major difference of our study is that, in most of previous studies, the network lifetime depends on the minimal nodal lifetime, i.e., the network depletes once a node in the network depletes, since the sensor nodes in those scenarios monitor different events; here, due to the correlations of measurements, CS enables us to care only about the number of activated sensor nodes. The network is considered depleted only when either there are not enough remaining sensor nodes to satisfy the cardinality requirement, or the remaining sensor nodes, with the sink nodes, can not form a connected graph. Thus, the flow approximation and the existing solutions cannot be directly applied in our setting. In this paper we adopt the concept of energy balancing and we characterize its relation to the maximum lifetime.

\subsection{Compressive sensing for data gathering}
\begin{figure}[t]
	\centering
	\includegraphics[width=3in]{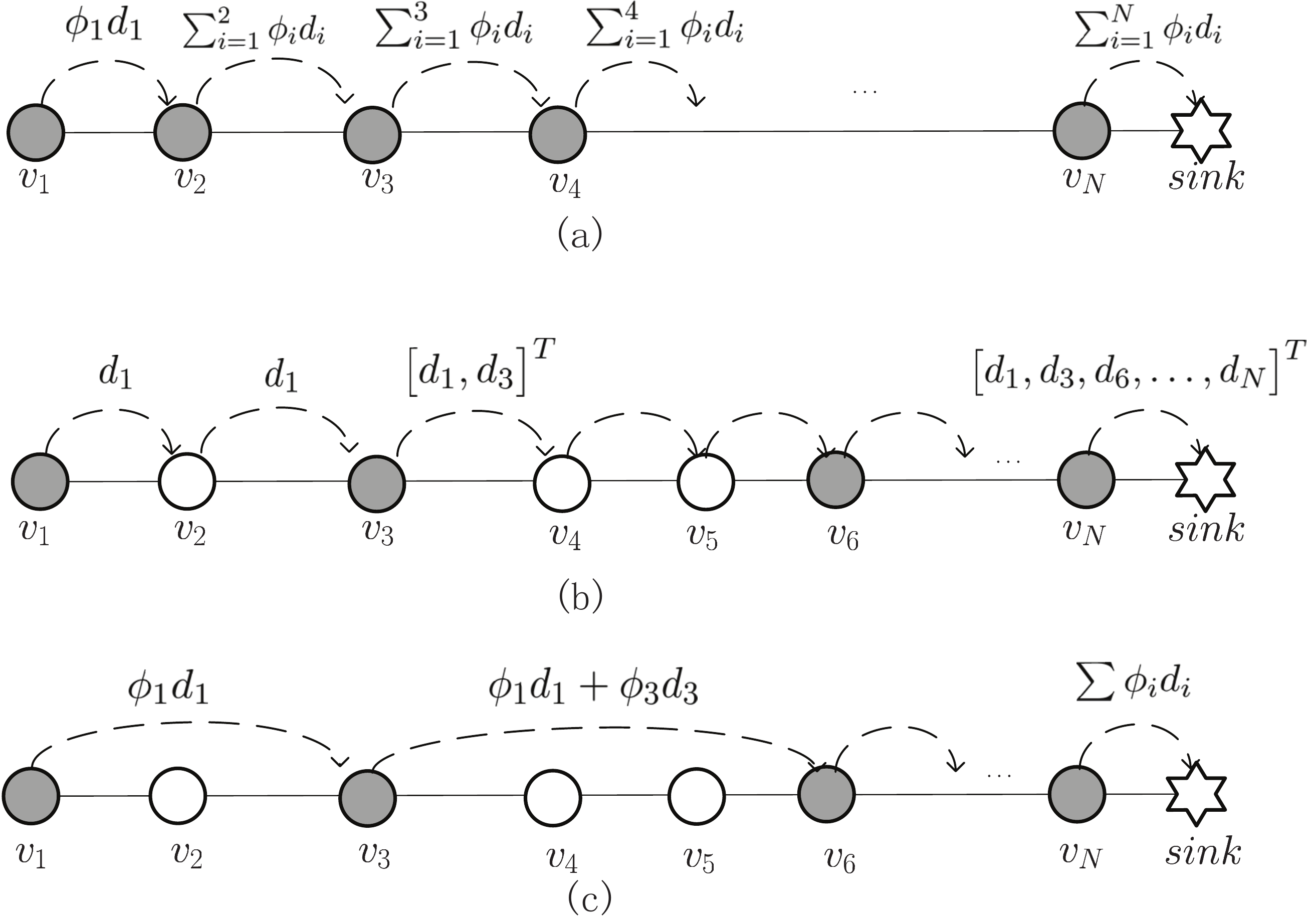}\\
	\vspace{-0.1cm}
	\caption{Data gathering in (a) Compressive Data Gathering (CDG) \cite{luo2009compressive}; (b) Compressed Sparse Function (CSF) \cite{xu2013efficient}; (c) Proposed scheme, where $d_i$ is the measurement of sensor node $v_i$, and $\vec{\phi}_i$ is a sampling vector with size $M\ll N$. In (a) and (c), every active node multiplies its measurement $d_i$ with vector $\vec{\phi}_i$, adds the result with the vector received from the previous active node, and then transmits the new vector to its next hop node. Because the active nodes transmit vectors with size $M$, they have the same size of payloads. }\label{CSexample}
	\vspace{-0.3cm}
\end{figure}
For the sake of completeness, we describe the operation of existing data gathering schemes based on CS. Suppose that the sink node in a network wants to collect the measurements of $N$ sensor nodes. In work \cite{luo2009compressive}, the proposed compressive data gathering (CDG) algorithm operates as shown in Fig.~\ref{CSexample} (a), in which each sensor node multiplies its local measurement $d_i\in\mathbb{R}$, for $i=1,\ldots,N$ with a random vector $\vec{\phi}_i$ of dimension $M\ll N$, adds the product $\vec{\phi}_id_i$ with the vector $\sum_{k=1}^{i-1}\vec{\phi}_id_i$ it receives from its neighbour, and then transmits the summation $\sum_{k=1}^i\vec{\phi}_id_i$ to its next-hop sensor node. At the sink node, the measurements of the sensor nodes, $\vec{y}=[\vec{\phi}_1,\ldots,\vec{\phi}_N]\vec{d}$, are recovered by solving an $l_1$-minimization problem as follow:
\begin{align*}
\vec{\hat{x}}=\arg\min_{\vec{x}} \|\vec{x}\|_{l_1}\quad \text{s.t. } \|\vec{y}-\vec{\Phi}\vec{\Psi}\vec{x}\|_{l_2}^2\leq \varepsilon\,,
\end{align*}
where $\vec{\Psi}$ is the basis on which $\vec{d}$ is sparse, and where $l_1$ and $l_2$ are the Manhattan norm and the Euclidean norm respectively. Then, $\vec{\hat{d}}=\vec{\Psi}\vec{\hat{x}}$.  In this case, every sensor node only transmits messages of size $O(M)$, which balances the energy consumptions of the sensor nodes and also reduces the overall data traffic. 

Fig.~\ref{CSexample} (b) shows the Compressed Sparse Function (CSF) algorithm \cite{xu2013efficient} for data gathering. In the figure, the grey sensor nodes sense and transmit their measurements to the next-hop sensor nodes, whereas the white ones act as relay nodes. As long as the sink node collects $M$ out of $N$ data measurements, the CSF algorithm can recover the remaining $N-M$ measurements. The idea is based on the mapping of the reading of the sensor nodes to their locations (or ids). Denote the mapping function $f(x)$ where $x$ is the location of the id of the sensor nodes. Since the sensor nodes are densely deployed, $f(x)$ could be represented as $f(x)=\vec{\Psi} (x)^T\vec{c}$, where $\vec{\Psi} (x)^T$ is a basis such as the type-IV DCT basis, and $\vec{c}$ is the sparse coefficient of $f(x)$ on the basis. Then, suppose the ids of the activated sensor nodes are $a_1,\ldots,a_M$, the sink node has the data of $\vec{y}=[f(a_1),\ldots,f(a_M)]^T=[\vec{\Psi}(a_1)^T,\ldots,\vec{\Psi}(a_M)]^T\vec{c}$. The sink node estimates $\vec{c}$ by the $l_1$-minimization problem similar to the CDG approach, and then the mapping function $f(x)$ is retrieved. Based on $f(x)$, the measurements of the white nodes are also retrieved.

Considering a dense network, in our previous work \cite{rong2015energy}, we proposed the data gathering scheme shown in Fig.~\ref{CSexample}~(c), where only the grey nodes are active and transmit data in the CDG way to the sink node. Since each active node transmits a calculated vector based on the summation of its measurement and its received vector, the packet sizes of the nodes are the same, and the energy consumption of the active nodes is balanced. The sink node first uses $l_1$-minimization to estimate the measurement of the grey sensor nodes, and then we use CSF to estimate the measurement of the white sensor nodes. The overall data traffic is further reduced, and the energy consumption of the grey sensor nodes is balanced. If the activation of the sensor nodes is decided carefully in each monitoring timeslot, the energy consumption of all the sensor nodes can be approximately balanced. However, understanding whether the network lifetime can be maximized by such an energy balancing approach is a fundamental open question.

Compared to our previous work \cite{rong2015JSAC}, the major difference of this paper is two-fold: 1) our previous work provides an efficient method to solve the energy balancing problem, whereas this paper characterizes the performance of energy balancing in terms of network lifetime; 2) in this paper, we achieve another upper bound of network lifetime by transforming the original problem into a maximum flow with cardinality constraint problem. This upper bound is tighter than the bound of \cite{rong2015JSAC} and provides useful insight for the structure of the problem.

\section{Problem Formulation and Preliminaries} \label{sec:problem}

We consider a WSN consisting of two tiers of nodes that monitors an area of line shape. Characteristic such examples are a pipeline in a water distribution network, a tunnel, or a bridge. The first tier consists of battery-powered sensor nodes that are densely deployed in the monitored area. Their role is to sense and relay data to a set of sink nodes. The second tier consists of 
sink nodes, which are grid-powered and are deployed at the two ends of the line. They collect data from the sensor nodes, and transmit the data to a remote monitoring center. Due to the length of the monitored area and the comparative small communication range of the sensor nodes, a multi-hop communication path from the sensor nodes to the sink node has to be established. 

Since battery replacement is not easy for the applications mentioned above, a main objective is to maximize the network lifetime. Intuitively, it is beneficial to keep alive as much of the sensor nodes as possible, which motivates the design of activation algorithms based on energy balancing, i.e., preferably activate the nodes with more residual energy. Thus, the major problem to be considered here, is whether the maximum network lifetime can be achieved by the energy balancing approach, or (if not), what is the performance of the energy balancing approach in terms of network lifetime.

In the following, we use a scenario of pipeline monitoring for water distribution network to better illustrate the necessary concepts. However, all the provided results hold for any linear network. The major notations are listed in Table~\ref{table:notation}.
\begin{table}\small	
	\centering
	\caption{Major notations used in the paper}\label{table:notation}
	\begin{tabular}{|c|l|}
		\hline symbols & meanings \\ 
		\hline $\mathcal{G}(\vec{x}(t))$ &  \tabincell{l}{induced graph of active sensor nodes \\  and sink nodes}\\
		\hline $\mathcal{N}(v_i)$ & set of neighbour nodes of $v_i$\\
		\hline $C_{ij}$ & capacity of arc $\langle v_i^{\rm{out}},v_j^{\rm{in}}\rangle$\\
		\hline $E_i$ & \tabincell{l}{ratio of nodal battery to nodal energy \\ consumption of $v_i$}\\		
		\hline $M_c$ & \tabincell{l}{minimum number of sensor nodes to be\\activated to satisfy connectivity constraint}\\
		\hline $M_{\rm{cs}}$ & \tabincell{l}{required number of active sensor nodes \\ in a timeslot due to compressive sensing}\\
		\hline $N$ & number of sensor nodes \\ 
		\hline $\bar{T}^f$ & \tabincell{l}{upper bound of lifetime achieved by\\ maximum flow approach}\\
		\hline $T_G$ & \tabincell{l}{lifetime achieved by energy balancing\\ approach}\\
		\hline $T_{\max}$ & maximum lifetime of the network\\
		\hline $\vec{Q}_i$ & \tabincell{l}{a column vector representing a feasible \\ activation profile}\\
		\hline $\vec{R}_i$ & \tabincell{l}{a route from $s_l$ to $s_r$, and is represented\\  by a list of nodes in the route}\\
		\hline $\vec{R}_i^+$ & set of forward arcs in  $\vec{R}$\\
		\hline $\vec{R}_i^-$ & set of backward arcs in $\vec{R}_i$\\
		\hline $p_i$ & normalized residual energy of $v_i$\\
		\hline $s_l$ and $v_0$ & the leftmost sink\\
		\hline $s_r$ and $v_{N+1}$ & the right most sink\\		
		\hline $u_{i,j,t}$ & flow from $v_i$ to $v_j$ at timeslot $t$\\
		\hline $v_i$ & sensor node $i$\\
		\hline $x_i(t)$ & activation of $v_i$ at timeslot $t$\\
		\hline $z_i$ & number of activations of $\vec{Q}_i$\\
		\hline
	\end{tabular} 
\end{table}

A WSN for monitoring a single pipeline can be represented by a communication graph $\mathcal{G}=(\mathcal{V},\mathcal{E})$, where vertex set $\mathcal{V}$ represents both the sensor nodes and the sink nodes, and edge set $\mathcal{E}$ represents the links among nodes. Suppose there are $N$ sensor nodes, then we denote $s_l$ the leftmost sink node, $s_r$ the rightmost sink node, and $v_1,v_2,\ldots,v_N$ the sensor nodes from left to right. For simplicity, sink nodes $s_l$ and $s_r$ are also represented as $v_0$ and $v_{N+1}$. Let $r_i$ be the transmission range of $v_i$, then $\langle v_i,v_j\rangle \in \mathcal{E}$ if and only if the distance between $v_i$ and $v_j$ is smaller than or equal to $r_i$. Also, we denote $\mathcal{N}(v_i)=\{v_j|\langle v_i,v_j\rangle\in\mathcal{E}\}$ the set of neighbours of $v_i$, and $\mathcal{N}_-(v_i)=\{v_j|v_j\in\mathcal{N}(v_i)\wedge j>i\}$ the downstream set of $v_i$. 
Our analysis relies on the following two assumptions that generally hold in water distribution networks\footnote{Those assumptions are necessary to derive the analytical results. More details on the general performance can be found in our previous work~\cite{rong2015JSAC}.}:
\begin{assumption}
	\label{assumption:transmissionrange}
	All sensor nodes and the sink nodes are characterized by the same communication range $r_i=r$.
\end{assumption}
\begin{remark}
	We suppose that the transmission power of the sensor nodes are fixed to a pre-set value. Since in a dense network, a node could have multiple neighbors for data relaying, the transmission power of a node could be set to the minimum value to save energy. Thus, the communication range of all the sensor nodes is the same. However, in the numerical simulations, we also examine the performance of Algorithm 1 when this assumption does not hold.
\end{remark}
\begin{assumption}
	\label{assumption:line}
	All sensor nodes and sink nodes in the same pipeline are deployed in a line.
\end{assumption}
\begin{remark}
	This assumption is reasonable as the diameter of the pipeline is small compared to the length of a pipeline. Thus, a pipeline sensor network can be considered as a line.
\end{remark}

For energy saving purposes, nodes can be deactivated. Time is slotted. In a timeslot $t$, a sensor node can either be activated to sense and transmit data, or be in the sleeping mode. The activated sensor nodes transmit data in the CDG way \cite{luo2009compressive}, i.e., every node transmits a vector of the same size based on the projection of its measured data and the vector it receives. Thus, the payloads of the transmitted packets at each active node are the same, and the energy consumption is approximately the same. Therefore, we may normalize the energy consumption for the active sensor nodes in a timeslot to 1 for simplicity. Then the energy budget of $v_i$, $E_i\in \mathcal{Z}^{+}$, is characterized as the number of timeslots that a node can be activated. Let $E_i(t)$ denote the number of timeslots that $v_i$ can be activated from timeslot $t$, which can be considered as the residual energy at $t$, and $E_i(1)=E_i$.

In the following, we first formulate the lifetime maximization problem. Then, we introduce an equivalent multi-dimensional knapsack problem, which allows us to solve the lifetime maximization problem for small networks. Last, we present the energy balancing problem and a solution algorithm, which are then used to study the achievable  performance of energy balancing in terms of network lifetime.

\subsection{Lifetime maximization problem}
We define the lifetime of a WSN to be the operating time until either WSN becomes disconnected, or the monitoring performance of the WSN cannot be guaranteed. In each timeslot, the connectivity and the monitoring performance requirement of the active sensor network must be satisfied. Let binary variable $x_i(t)$ indicate whether node $v_i$ is active at timeslot $t$. 
Then, the energy dynamics of $v_i$ can be written as $E_i(t+1)=E_i(t)-x_i(t)$, and the scheduling problem considered in this paper is to determine $\vec{x}(t)=[x_1(t),\ldots,x_N(t)]^T,\forall t$.

Let $\mathcal{G}(\vec{x}(t))$ denote the induced graph of active sensor nodes and the sink nodes. Then, the connectivity constraint is defined as follows:
\begin{definition}\label{def:connectivity}
	\textbf{(Connectivity Constraint)} The activation of the sensor nodes $\vec{x}(t)$ satisfies the connectivity constraint if and only if the induced graph $\mathcal{G}(\vec{x}(t))$ is connected.
\end{definition}
\begin{remark}
	To check the connectivity of a subgraph, one approach could be breadth-first search. However, since the connectivity checking is not the major objective of the paper, we do not elaborate further. More details can be found in Chapter 3.2 of~\cite{kleinberg2006algorithm}.
\end{remark}

Regarding monitoring performance, our previous works \cite{rong2015energy,rong2015JSAC} have shown that the estimation error by CS is related to the number of active nodes, $m$, where $m$ is much smaller than the number of sensor nodes $N$, i.e., $m\ll N$. Thus, the requirement on monitoring performance can be specified as a cardinality constraint defined as follows:

\begin{definition}\label{def:cardinality}
	\textbf{(Cardinality Constraint)} The activation of the sensor nodes $\vec{x}(t)$ satisfies the cardinality constraint if and only if $\sum x_i(t)\geq M_{\rm{cs}}$, where $M_{\rm{cs}}$ is determined by the required estimation error of the measured data.
\end{definition}

Then, the lifetime maximization problem can be formulated as an optimal control problem as follows:
\begin{subequations}
	\label{ProblemOPC}
	\begin{align}
	&\max_{\vec{x}(t),t=1,\ldots,T}\quad \sum_{t=1}^T 1 \label{ProblemOPC:objective}\\
	\text{s.t.}\quad & E_i(t+1)=E_i(t)-x_i(t),\forall i,1\leq t\leq T,\label{ProblemOPC:dynamic}\\
	& E_i(1)=E_i,\forall i,\label{ProblemOPC:initstate}\\
	& E_i(t)\geq 0,\forall i,1\leq t\leq T+1,\label{ProblemOPC:energyconstraint}\\
	&\mathcal{G}(\vec{x}(t))\text{ is connected },\forall t,\label{ProblemOPC:connect}\\
	&\sum_{v_i\in\mathcal{V}}x_i(t)\geq M_{\rm{cs}},\forall  1\leq t\leq T,\label{ProblemOPC:cardinality}\\
	&x_i(t)\in \{0,1\},\forall i,t\label{ProblemOPC:binary}\,,
 	\end{align}
\end{subequations}
where Constraint~\eqref{ProblemOPC:dynamic} is the dynamic of the energy of the sensor nodes, Constraint~\eqref{ProblemOPC:initstate} is the initial state of the WSN, \eqref{ProblemOPC:energyconstraint} is the non-negative constraint on the energy of sensor nodes, \eqref{ProblemOPC:connect} is the connectivity constraint, \eqref{ProblemOPC:cardinality} is the cardinality constraint. Also, we have that Constraints~\eqref{ProblemOPC:dynamic}-\eqref{ProblemOPC:energyconstraint} can be equivalently captured by the following energy budget constraint:
\begin{align}
\sum_{t=1}^Tx_i(t)\leq E_i,\forall i.\label{Problem0:Energy}\,,
\end{align}
where the proof is in Appendix~\ref{Appendix:EnergyBudget}.

Notice that, given a feasible $\vec{x}(t),\forall t$ that satisfies Constraints~\eqref{Problem0:Energy}, and ~\eqref{ProblemOPC:binary}, we can construct $E_i(1)=E_i$, and $E_i(t+1)=E_i(t)-x_i(t)$, such that $E_i(t)\geq 0,\forall i,1\leq t\leq T+1$ always holds. Thus, indeed we can replace Constraints~\eqref{ProblemOPC:dynamic}-\eqref{ProblemOPC:energyconstraint} by Constraint~\eqref{Problem0:Energy}. 
This optimization problem  is particularly challenging due to the binary nature of activation variables and cardinality constraint (\ref{ProblemOPC:cardinality}), as we articulate below.

\begin{remark}
	\label{rmk:connectivity}
	Note that Problem~\eqref{ProblemOPC} addresses only which set of nodes should be activated, and not how the routing of measurements to the sink in a multi-hop fashion should be performed. Under Assumptions~\ref{assumption:transmissionrange} and~\ref{assumption:line}, all vertices in a connected subgraph of $\mathcal{G}$ can form a connected routing path, i.e., they are in a line topology and every node receives data from at most one node and transmits data to at most one node (see more details from the proof Theorem 1 in our previous work~\cite{rong2015JSAC}). Thus, it is guaranteed that a forward to the nearest active neighbour routing is always an optimal routing strategy. In other words, the induced graph of the active sensor nodes and the sinks is a connected graph, if and only if there is a path in the induced graph from $v_0$ to $v_{N+1}$ that passes only through all the active nodes. Thus, it is equivalent to replace Constraint~\eqref{ProblemOPC:connect} by the requirement of the existence of path from $v_0$ to $v_{N+1}$ for any $t$ without changing the optimal solution of the original problem, as we will do in Section~\ref{sec:performance} to simplify analysis.
\end{remark}

\subsection{Knapsack approximation for small WSNs}\label{sec:knapsack}
Lifetime maximization Problem (1) is NP-Hard~\cite{rong2015JSAC}. In order to provide insight on the complexity and the structure of the problem, we show that it can be cast as a knapsack optimization problem, for which a solution method is known. However, the method is practical only for small networks. We also use it in the numerical evaluation part as a benchmark.


%

To begin with, we define activation profile as follow:
\begin{definition}\label{def:activation profile}
		\textbf{(Activation Profile)} An activation profile is a group of sensor nodes that satisfies the connectivity constraint. We say an activation profile is feasible if and only if it also satisfies the cardinality constraint.
\end{definition}
 Then, we may equivalently reformulate the maximum lifetime problem as stated in the following lemma:

\begin{lemma}
	\label{prop:ILP}
	Consider a WSN $\mathcal{G}$. Let $L$ be the total number of feasible activation profiles. Each feasible activation profile is represented by a column vector $\vec{Q}_l=[q_l(1),\ldots,q_l(N)]^T$. Here,  $q_l(i)=1$ if and only if sensor node $v_i$ belongs to profile $l$, otherwise $q_l(i)=0$. Define vector $\vec{z}=[z_1,\ldots,z_L]^T$, where $z_i$ denotes the number of timeslots that profile $i$ is chosen for activation. Then, the lifetime maximization Problem~(\ref{ProblemOPC}) is equivalent to the following problem:
	\begin{subequations}
		\label{problem:ILP}
		\begin{align}
		T=\max_{\vec{z}}\quad & \sum_{l=1}^Lz_l\\
		\text{s.t.}\quad & \vec{Q}\vec{z}\leq \vec{E},\label{problem:ILP:constraint1}\\
		&z_l\in \mathcal{Z}^+,\forall l\in\{1,\ldots,L\}\,,
		\end{align}
	\end{subequations}
	where $\vec{Q}=[\vec{Q}_1,\ldots,\vec{Q}_L]$ is the set of all feasible profiles, $\vec{E}=[E_1,\ldots,E_N]^T$ is the vector of initial energies of all sensor nodes, and $\mathcal{Z}^+$ is the set of non-negative integers.
\end{lemma}
\begin{IEEEproof}
	The proof is in Appendix~\ref{Appendix:MDK}.
\end{IEEEproof}

Lemma~\ref{prop:ILP} shows that the maximum lifetime problem under cardinality and connectivity constraints can be turned into a multi-dimensional knapsack (MDK) problem \cite{bertsimas2002approximate}. In our case, the knapsack corresponds to the energy budget of each sensor nodes, and each profile corresponds to an item of value $1$.

\begin{remark} There are several methods to solve MDK problems, such as branch-and-bound, dynamic programming, and heuristic algorithms. We compare them in terms of complexity and optimality in Table~\ref{table:MDKcompare}. Branch-and-bound, and dynamic programming can achieve optimal solution, however, the complexity of branch-and-bound is as high as exhaustive search in the worst case, whereas dynamic programming suffers from the curse of dimensionality and is not a scalable approach. Therefore, they cannot be applied in our scenario, where network size is large and nodes have limited computational and storage capability. However, we use branch-and-bound algorithm in  simulation part for comparison purposes. Besides, heuristic methods such as genetic networks, even though with low complexity, cannot guarantee optimality. Thus, the transformation into an MDK problem is only valid for small scale network instances. For large scale (dense) networks, we consider approximating Problem~\eqref{ProblemOPC} by an energy balancing problem described in the next subsection.
\end{remark}

\begin{table}\small
	\caption{Comparison of different methods for multi-dimensional knapsack problem}
	\centering
	\label{table:MDKcompare}
	\begin{tabular}{lll}
		\hline
		\multicolumn{1}{c|}{Methods} & \multicolumn{1}{c}{Complexity}   & \multicolumn{1}{c}{Optimality}  \\ \hline
		\multicolumn{1}{c|}{Branch-and-Bound}  & \multicolumn{1}{c}{Unknown} & \multicolumn{1}{c}{Yes} \\
			\multicolumn{1}{c|}{Dynamic Programming}  & \multicolumn{1}{c}{Curse of dimension} & \multicolumn{1}{c}{Yes}   \\
			 \multicolumn{1}{c|}{Heuristic}  & \multicolumn{1}{c}{Low} & \multicolumn{1}{c}{Not sure}   \\
			 \hline
	\end{tabular}
	\vspace{-0.3cm}
\end{table}
\subsection{Energy balancing problem}
In our previous work \cite{rong2015energy}, we proposed an energy balancing problem, together with a solution method. Since in this paper we investigate the fundamental properties of the energy balancing problem from the point of view of network lifetime maximization, we give the necessary details in the following:




Recall that $E_i(t)$ is the residual energy of $v_i$ at timeslot $t$, we define its normalized residual energy as $p_i(t)=E_i(t)/E_i$. We denote $\mathcal{V}(t)=\{v_i\in \mathcal{V}|E_i(t)\geq 1\}$ the set of candidate sensor nodes to be potentially activated at timeslot $t$. Then, we posed the energy balancing problem as a sequence of problems, and one for each timeslot $t$ as follows:
\begin{subequations}
	\label{SubProblem0}
	\begin{align}
		&\max_{\vec{x}} \sum_{i\in \mathcal{V}} x_{i}p_{i}\\
		\text{s.t.}\quad & \sum_{v_i\in \mathcal{V}} x_{i}=\max\{M_{\rm{cs}},M_c\}, \label{eq:sub0_conm}\\
		& \mathcal{G}(\vec{x}) \text{ is connected},\label{eq:sub0_connection}\\
		& x_i\in\{0,1\},\; \forall i\in\mathcal{V},
	\end{align}
\end{subequations}
where time index $t$ is discarded for notational simplicity, and $M_c$ is the minimum number of sensor nodes that must be activated to satisfy the connectivity constraint. 

The intuition behind the energy balancing approach is that, in each timeslot, the number of active sensor nodes has to be as small as possible. Out of the feasible activation profiles with the same number of active sensor nodes, the profile with maximum normalized residual energy is preferred.
Recall that $M_{\rm{cs}}$ is the minimum number of sensor nodes that should be activated to ensure monitoring performance.  When $M_{\rm{cs}}<M_c$, one cannot find a route from $s_l$ to $s_r$ which passes exactly $M_{\rm{cs}}$ sensor nodes. In this case, the requirement of $\sum_{v_i\in\mathcal{V}}x_i=M_{\rm{cs}}$ contradicts to Constraint~\eqref{eq:sub0_connection}. It gives us that $\sum_{v_i\in\mathcal{V}}x_i$ should be $M_c$ instead.
Thus, even though Constraint~\eqref{eq:sub0_connection} is given, the $M_c$ in Constraint~\eqref{eq:sub0_conm} is not redundant. 

	\begin{algorithm}[t]\small
		\caption{Optimal activation schedule for Problem (\ref{SubProblem0}) \cite{rong2015JSAC}}
		\begin{algorithmic}[1]
			\label{SubProblemSolver}
			\REQUIRE Adjacency matrix $\vec{A}$, the minimum number of active node $M_{cs}$, and the normalized residual energy of the sensor nodes $\vec{p}$.
			\ENSURE  A set of sensor nodes $\mathcal{V}_A$ that need to be activated.
			\STATE Find the minimum number of sensor nodes, $M_c$, that satisfy the connectivity constraint
			\IF {$M_c< +\infty$} 		
			\item[] \COMMENT{Find the minimum number of nodes that satisfy both connectivity constraint and cardinality constraint}
			\STATE $m=\max\{M_c,M_{\rm{cs}}\}$
			\item[] \COMMENT{Solve Problem (\ref{SubProblem0}) by dynamic programming}
			\STATE Calculate $g(s_l,m)$ according to Eq. (\ref{iteration0}) and the corresponding $\mathcal{V}_A$  
			\RETURN $\mathcal{V}_A$
			\ELSE 
			\RETURN $\emptyset$
			\ENDIF
		\end{algorithmic}
	\end{algorithm}
	
		\begin{algorithm}[t]\small
			\caption{Activation based on Energy Balancing~\cite{rong2015JSAC}}
			\begin{algorithmic}[1]
				\label{Alg:solver}
				\REQUIRE Adjacency matrix $\vec{A}$, the minimum number of active node $M_{cs}$, the battery of the nodes $E_i,\forall i$
				\ENSURE  Network lifetime $T$
				\STATE Set $t\leftarrow1$, $Flag\leftarrow TRUE$, $E_i(t)=E_i$
				\WHILE{$Flag$}
				\STATE Set $p_i\leftarrow E_i(t)/E_i$, $\vec{p}=\{p_1,\ldots,p_N\}$
				\item[] \COMMENT{Node activation based on Algorithm~\ref{SubProblemSolver}}
				\STATE Find $M_c$ for the connectivity constraint
				\STATE $\mathcal{V}_A\leftarrow$ Call Algorithm~\ref{SubProblemSolver} with input $\vec{A},M_{cs},\vec{p}$
				\IF{$\mathcal{V}_A\neq \emptyset$}
				\STATE Set $t\leftarrow t+1$, $E_j(t)\leftarrow E_j(t-1)-1,\forall j\in\mathcal{V}_A$
				\ELSE
				\STATE Set $Flag\leftarrow FALSE$
				\ENDIF
				\ENDWHILE
				\RETURN $T\leftarrow t-1$
			\end{algorithmic}
		\end{algorithm}

In \cite{rong2015energy}, we developed an algorithm to solve Problem (\ref{SubProblem0}). The details of the procedure are shown in Algorithm~\ref{SubProblemSolver}. 
First, Algorithm~\ref{SubProblemSolver} finds $M_c$ by a shortest path algorithm such as Dijkstra's algorithm in Line 1, namely finds the shortest path from $v_0$ to $v_{N+1}$, where the weights of all the edges are 1. Then, the minimum number of sensor nodes, $m$, that satisfies both the connectivity and the cardinality constraints is calculated in Line 3. Knowing $m$, we can solve Problem (\ref{SubProblem0}) by dynamic programming (Line 4), where $g(v_i,k)$ represents the maximum sum of normalized residual energy of $k$ connected sensor nodes among $v_{i+1}$ to $v_N$, and it is calculated as
\begin{equation}
	\label{iteration0}
	g(v_i,k)=\begin{cases}
		\max\limits_{v_j\in \mathcal{N}'_{-}(v_i)}\{g(v_j,k{-}1){+}p_j\} & \text{if } k{>}0, \mathcal{N}'_{-}(v_i) {\neq} \emptyset\\
		0 & \text{if } k{=}0, s_r{\in}\mathcal{N}_-(v_i)\\
		-\infty & \text{otherwise}\,,
	\end{cases}
\end{equation}
where $p_j$ is the normalized residual energy of sensor node $v_j$. Recall that $\mathcal{N}_-(v_i)=\{v_j|v_j\in\mathcal{N}(v_i)\wedge j>i\}$ is the downstream set of $v_i$, $\mathcal{N}'_{-}(v_i)=\mathcal{N}_{-}(v_i)\backslash\{s_r\}$ is the set of sensor nodes in the downstream set of $v_i$. Notice that $g(v_i,k)$ is directly related to $g(v_j,k-1)$, where $v_j$ is in the neighbour of $v_i$. Thus, the nodes determined by this dynamic programming are connected. 

In our previous work \cite{rong2015JSAC}, we proposed to activate the sensor nodes as suggested by the solution of the energy balancing problem~\eqref{SubProblem0} in each timeslot. Then update the nodal normalized residual energy to be the input of the energy balancing problem in the next timeslot, until the problem is infeasible, as described by Algorithm~\ref{Alg:solver}. However, whether this approach could lead to the maximum network lifetime has not been analyzed before. Thus, \textit{the investigation of the fundamental properties of energy balancing in terms of network lifetime with a cardinality constraint} is the core contribution of this paper.

\section{Performance Analysis of the Optimal Activation Schedule}\label{sec:performance}
In this section, we analyze the performance of Algorithm~\ref{Alg:solver} in terms of network lifetime. We transform the lifetime maximization Problem~\eqref{ProblemOPC} to a maximum flow problem with a typical vertex capacity constraints 
and a new cardinality constraint. Such transformation  enables a better understanding of  the maximum lifetime problem, provides us a new lifetime upper bound, and also enables us to derive the performance bound of the proposed Algorithm~\ref{Alg:solver}.

\subsection{Lifetime maximization as a maximum flow problem}

The maximum lifetime Problem~\eqref{ProblemOPC} is a maximum flow problem with vertex capacities (see \cite{Ahuja1993network} for a description of these problems) with an additional  cardinality constraint. Let $u_{i,j,t}$ denote the flow from node $v_i$ to $v_j$ in timeslot $t$.
Then, the maximum flow with vertex capacity and cardinality constraints is formulated as follows:
\begin{subequations}
	\label{ProblemMFCC}
	\begin{align}
	& \max_{\vec{u}} \quad  T\\
	\text{s.t.} \quad  
	 &{{\sum_{v_j\in\mathcal{N}(v_i)}}{u_{i,j,t}}}{-}{\sum_{v_j:v_i\in\mathcal{N}(v_j)}}{u_{j,i,t}}\nonumber\\&\quad{=}\begin{cases}
	1, & i=0,\forall t=1,\ldots,T\\	
	-1,& i=N+1,\forall t=1,\ldots,T\\
	0, & \forall 1\leq i\leq N,\forall t=1,\ldots,T
	\end{cases},\label{ProblemMFCC:Connectivity}\\	
	&\sum_{v_j\in\mathcal{N}(v_i)}u_{i,j,t}\leq 1,\forall i,t\label{ProblemMFCC:noCycle}\\
	&\sum_{t=1}^T\sum_{v_j\in\mathcal{N}(v_i)}u_{i,j,t}\leq E_i,\forall i=1,\ldots,N,\label{ProblemMFCC:Budget}\\
	& \sum_{v_i\in\mathcal{V}}\sum_{v_j\in\mathcal{N}(v_i)}u_{i,j,t}\geq M_{\rm{cs}}+1, \forall t=1,\ldots,T,\label{ProblemMFCC:Cardinality}\\
	& u_{i,j,t}\in\{0,1\},\forall i,j,t\label{ProblemMFCC:binary}\,,
	\end{align}
\end{subequations}

Then, we have the following lemma:
\begin{lemma}
	\label{lemma:equivalentMFCC}
	Under Assumptions~\ref{assumption:transmissionrange} and~\ref{assumption:line}, Problem~\eqref{ProblemOPC} is equivalent to Problem~\eqref{ProblemMFCC}.
\end{lemma}
\begin{IEEEproof}
	The proof is in Appendix~\ref{Appendix:equivalentMFCC}.
\end{IEEEproof}


Problem~\eqref{ProblemMFCC} is a binary programming problem. If Constraint~\eqref{ProblemMFCC:binary} is relaxed to $0\leq u_{i,j,k}\leq 1,\forall i,j,k$, the problem becomes a linear programming problem. Then, an upper bound of WSN lifetime for Problem~\eqref{ProblemOPC} can be established, as stated by the following theorem:
\begin{theorem}
	\label{thm:lifetimeupperboundflow}
	Consider optimization Problem~\eqref{ProblemOPC} for a WSN that follows Assumptions~\ref{assumption:transmissionrange} and~\ref{assumption:line}. The WSN lifetime is upper bounded by $\bar{T}^f$, where $\bar{T}^f$ is the optimal value of Problem~\eqref{ProblemMFCC} with Constraint~\eqref{ProblemMFCC:binary} relaxed as $0\leq u_{i,j,t}\leq 1,\forall i,j,t$.
\end{theorem}
\begin{IEEEproof}
Suppose $T^{*}$ is the optimal value of Problem~\eqref{ProblemOPC}, then it is also the optimal value of Problem~\eqref{ProblemMFCC} according to Lemma~\ref{lemma:equivalentMFCC}. Further, as $\bar{T}^f$ is the optimal value of the relaxed Problem~\eqref{ProblemMFCC}, we have $T^{*}\leq \bar{T}^f$ which completes the proof.
\end{IEEEproof}
\begin{remark}
	This bound is quite tight if Assumptions~\ref{assumption:transmissionrange} and~\ref{assumption:line} hold, as will be shown in Section V. Notice that the relaxed Problem~\eqref{ProblemMFCC} is a linear optimization problem and solvable. Consequently, we can use a bisection approach to find $\bar{T}^f$, as discussed in Appendix~\ref{Appendix:RelaxMFCC}. Besides, one may derive a good solution by rounding the result of the relaxed Problem~\eqref{ProblemMFCC}\footnote{However, the main difficulty is to determine the rules of rounding such that the result satisfies both connectivity and cardinality constraints, and leads to the maximum network lifetime. This is left as a future work.}.  
\end{remark}

Based on Theorem~\ref{thm:lifetimeupperboundflow}, we study the performance of Algorithm~\ref{Alg:solver} from the perspective of maximum flow problem. 
We first turn the maximum flow Problem~\eqref{ProblemMFCC} with vertex capacities to a maximum flow problem with edge capacities according to the following remark.
\begin{remark}
	Problem~\eqref{ProblemOPC} can be formulated as a maximum flow problem with edge capacities~\cite{Como2015Throughput} and cardinality constraints. The basic idea is to substitute each node $v_i$ with two nodes $v_i^{\rm{in}}$ and $v_i^{\rm{out}}$ connected by an arc $\langle v_i^{\rm{in}},v_i^{\rm{out}}\rangle$ with capacity $E_i$. More details can be found in Appendix~\ref{Appendix:EdgeCapacity}.
\end{remark}


Then, we show how the problem can be solved via a modified maximum flow algorithm. For such a purpose, we introduce some additional notations.  Let $f_{ii}$ be the flow on arc $\langle v_{i}^{\text{in}},v_{i}^{\text{out}}\rangle$, and $f_{ij}$ the flow on arc  $\langle v_{i}^{\text{out}},v_{j}^{\text{in}}\rangle$. The capacity of arc $\langle v_{i}^{\text{in}},v_{i}^{\text{out}}\rangle$ is $C_{ii}=E_i$, and the capacity of arc $\langle v_{i}^{\text{out}},v_{j}^{\text{in}}\rangle$ is $C_{ij}=+\infty$, as shown in Fig.~\ref{fig:vertexcap}. Given a route $\vec{R}_a=\langle s_l, v_{a_1},\ldots,v_{a_k}, s_r\rangle$, we say that arc $\langle v_{a_i},v_{a_{i+1}}\rangle$ belongs to the set of forward arcs of $\vec{R}_a$, which is denoted by $\vec{R}_a^+$, if and only if $\langle v_{a_i},v_{a_{i+1}}\rangle \in \mathcal{E}'$. Otherwise, the arc belongs to the set of backward arcs of $\vec{R}_a$, which is denoted by $\vec{R}_a^-$.
Then the maximum flow increment of $\vec{R}_a$ is defined as $\delta_a=\min\{\{C_{a_ia_j}-f_{a_ia_j}|\langle v_{a_i},v_{a_j}\rangle \in \vec{R}_a^+\},\{f_{a_ia_j}|\langle v_{a_i},v_{a_j}\rangle \in \vec{R}_a^-\}\}$. $\vec{R}_a$ is said to be unblocked if and only if $\delta_a>0$. Then we can use a modified Ford-Fulkerson Algorithm to find which nodes should  be activated at each timeslot, so that the corresponding route at each timeslot is feasible.

\subsection{A modified maximum flow algorithm based on Ford-Fulkerson Algorithm}
The derived modified Ford-Fuklerson Algorithm works as follows. We find an unblocked route $\vec{R}_i$ from $s_l$ to $s_r$ in $\mathcal{G}'$ that contains no backward arcs and  passes at least $M_{\rm{cs}}$ arcs with capacity less than $+\infty$. This is equivalent to passing at least $M_{\rm{cs}}$ sensor nodes in $\mathcal{G}$, and hence it corresponds to one route in $\vec{R}$. We perform an augmentation along route $R_i$ with increment 1, i.e., the flows $f$ of all the arcs in the route $R_i$ increase by 1. Then, we find an unblocked route that contains no backward arcs until there is no such unblocked route in $\mathcal{G}'$ again. This operation gives a sequence of routes $\vec{R}(1),\vec{R}(2),\ldots,\vec{R}(T)$. If we are unable to find an unblocked route that contains backward arcs at $T+1$, then activation scheme $\vec{R}=[\vec{R}(1),\vec{R}(2),\ldots,\vec{R}(T)]$ leads to maximum network lifetime.


Notice that this algorithm does not allow choosing an unblocked route with backward arcs in each timeslot, it requires an exhaustive search, and thus is not practical. However, it suggests the following lemma:
\begin{lemma}
	\label{lemma:suboptimal}
	Consider a WSN satisfying Assumptions~\ref{assumption:transmissionrange} and~\ref{assumption:line}. If an unblocked route with backward arcs exists when Algorithm~\ref{Alg:solver} terminates, we can alter one of the previous activation decisions to extend network lifetime by 1.
\end{lemma}
\begin{IEEEproof}
	The proof is in Appendix~\ref{Appendix:suboptimal}.
	\end{IEEEproof}
This lemma shows that \textit{the existence of an unblocked route with backward arcs suggests the suboptimality of an activation}. Furthermore, extending the network lifetime in such way requires an unblocked route  with backward arcs ($\vec{R}_b$ in the proof) and an unblocked route with no backward arcs ($\vec{R}_a$ in the proof) that we have selected before the WSN expires. This gives us the worst case approximation ratio of the lifetime achieved by Algorithm~\ref{Alg:solver} to the maximum network lifetime, as will be shown in the next subsection.


\subsection{Performance analysis of Algorithm~\ref{Alg:solver}}
According to Lemma~\ref{lemma:suboptimal},
the more the unblocked routes we can find, the more suboptimal the activation is. Thus, we can analyze the gap of lifetime we achieve by Algorithm~\ref{Alg:solver} to the optimal lifetime value, by counting how many unblocked backward routes can be found when Algorithm~\ref{Alg:solver} terminates.

Then, the performance of Algorithm~\ref{Alg:solver} can be characterized by the following lemmas.

\begin{lemma}\label{lemma:backwardarc}
	Consider a WSN $\mathcal{G}$ that satisfies Assumptions \ref{assumption:transmissionrange} and \ref{assumption:line}. Algorithm~\ref{SubProblemSolver} is applied to determine the activation of the sensor nodes in each time slot. Let the achieved WSN lifetime be $t_1$, i.e., on $t_1+1$, no unblocked routes that contain only forward arcs can be found. If an unblocked route with backward arcs, $\vec{R}_b$, can be found at $t_1+1$, 
	then $\vec{R}_b$ does not contain any backward arc $\langle v_i^{\text{out}},v_i^{\text{in}}\rangle$ for any $i$.
\end{lemma}
\begin{IEEEproof}
	The proof is in Appendix~\ref{Appendix:backwardarc}.
	\end{IEEEproof}

\begin{figure}[t]
	\centering
	\includegraphics[width=0.4\textwidth]{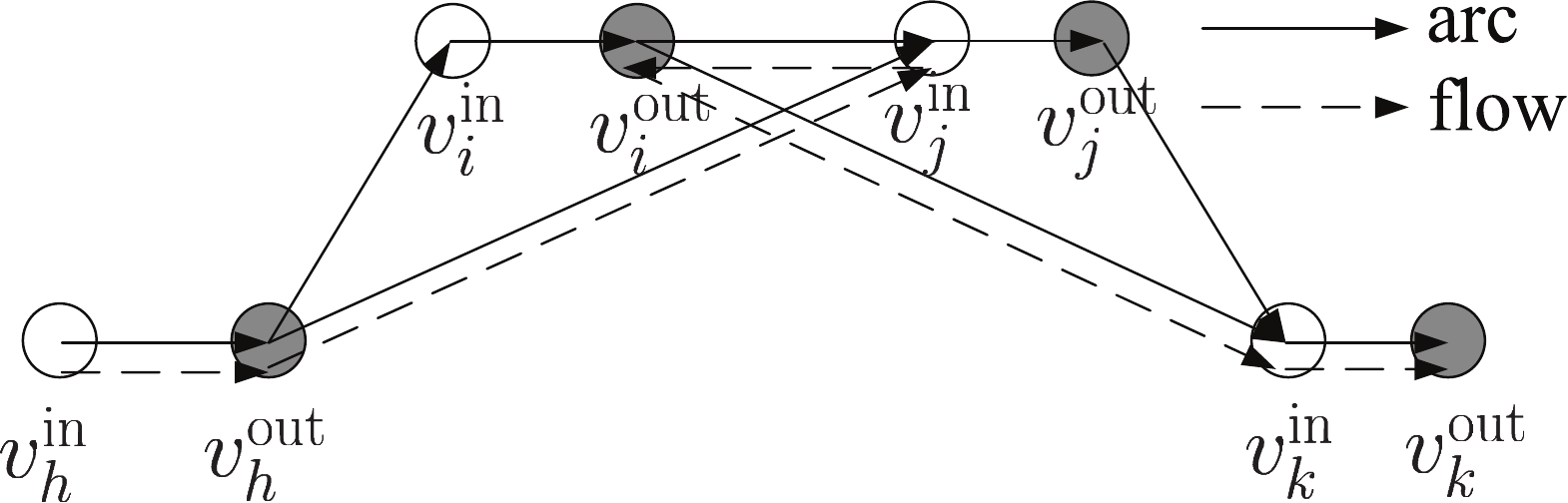}\\
	\caption{A subnetwork that contains a backward arc $\langle v_j^{\text{in}},v_i^{\text{out}}\rangle$}
	\label{fig:lemma}
	\vspace{-0.1cm}
\end{figure}

According to Lemma \ref{lemma:backwardarc}, if one can find unblocked routes with backward arcs at the end, the backward arcs should be $\langle v_j^{\text{in}}, v_i^{\text{out}}\rangle$, $j>i$. Thus, even though the route may contain several backward arcs, we can divide the route into several separated parts, each of which contains only one backward arc for easier analysis. Thus, we just need to focus on one of them as shown in Fig. \ref{fig:lemma}.


\begin{lemma}\label{lemma:1gap}
	Consider a WSN $\mathcal{G}$ that satisfies Assumptions \ref{assumption:transmissionrange} and \ref{assumption:line}. Algorithm~\ref{SubProblemSolver} is applied to determine nodes to activate in each timeslot. If a backward arc $\langle v_{j}^{\text{in}},v_{i}^{\text{out}}\rangle$ $(j>i)$ exists in an unblocked route $\vec{R}_b$ when the WSN lifetime expires, then the maximum flow increment of $\langle v_{j}^{\text{in}},v_{i}^{\text{out}}\rangle$ is 1.
\end{lemma}
\begin{IEEEproof}
	The proof is in Appendix~\ref{Appendix:1gap}.
	\end{IEEEproof}


From the proof of Lemma \ref{lemma:1gap}, we know that Algorithm~\ref{Alg:solver} could be suboptimal due to the existence of unblocked routes with backward arcs. However, the maximum flow increment of the unblocked route with backward arcs does not increase when we multiply the $E_i,\forall i$  with a positive integer $\eta$.  Then we have the following core result:
\begin{theorem}\label{prop:multiply}
	Consider a WSN that satisfies Assumptions \ref{assumption:transmissionrange} and \ref{assumption:line}, with initial energy $\vec{E}$. Let $T_{\max}(\eta)$ and $T_G(\eta)$ be the maximum network lifetime by Problem~\eqref{ProblemOPC} and the network lifetime achieved by Algorithm~\ref{Alg:solver}, with initial energy $\eta\vec{E}$. Then
	\begin{align*}
		\lim_{\eta\rightarrow +\infty}\frac{T_{\max}(\eta)-T_G(\eta)}{T_{\max}(\eta)}=0\,.
	\end{align*}
\end{theorem}

\begin{IEEEproof}
	$T_{\max}(\eta)-T_G(\eta)$ is bounded by the number of unblocked routes with backward arcs when the network expires. According to Lemma \ref{lemma:1gap}, this number does not increase if the $E_i,\forall i$ are multiplied by a positive integer. Thus, $T_{\max}(\eta)-T_G(\eta)$ is bounded. However, we know that $T_{\max}(\eta){\geq} \lfloor\eta\rfloor T_{\max}(1)$, and it tends to $\infty$ as $\eta$ tends to $\infty$. Thus,
	\begin{align*}
		\lim_{\eta\rightarrow +\infty}\frac{T_{\max}(\eta)-T_G(\eta)}{T_{\max}(\eta)}=0\,.
	\end{align*}	
	This concludes the proof.
\end{IEEEproof}

Furthermore, based on Lemma~\ref{lemma:suboptimal}, we have the approximation ratio of Algorithm~\ref{Alg:solver} as shown in the following theorem.
\begin{theorem}
	\label{thm:approxratio}
	Consider a WSN that satisfies Assumptions~\ref{assumption:transmissionrange} and~\ref{assumption:line}. Let $T_{\max}$ and $T_G$ be the maximum network lifetime and the lifetime achieved by Algorithm~\ref{Alg:solver}. Then, we have $T_G\geq 0.5T_{\max}$.
\end{theorem}
\begin{IEEEproof}
	 According to Lemma~\ref{lemma:suboptimal}, to extend the network lifetime requires us selecting an unblocked route with backward arcs and an unblocked route with no backward arcs  before the WSN expires. We call such pair of routes an incremental pair. Furthermore, we know from Lemma~\ref{lemma:1gap} that the maximum flow increment of each backward arc is at most 1. Thus, when Algorithm~\ref{Alg:solver} terminates, we have that,	 
	$T_{G}$, i.e., the summation of flows in unblocked routes without backward arcs, is larger or equal to the number of incremental pairs. Such number of incremental pairs is larger or equal to the additional network lifetime that can be extended by the backward arcs, which is $T_{\max}-T_G$. This gives us $T_G\geq 0.5T_{\max}$ and completes the proof.
\end{IEEEproof}
\begin{remark}
	The provided approximation ratio is tight. An example for $T_G=0.5T_{\max}$ can be shown using the topology of Fig.~\ref{fig:lemma}, where the left sink node is connected to $v_h$ and $v_i$ and the right sink node is connected to $v_j$ and $v_k$, and the cardinality constraint requires us to pick $2$ sensor nodes to activate in each time slot. The initial energy of $v_i,v_j,v_h,v_k$ are all 1. Then, it is easy to achieve that the maximum network lifetime is $2$, i.e., to activate $v_i$ and $v_k$ in one timeslot and to activate $v_h$ and $v_j$ in the other timeslot. In this case, the lifetime achieved by Algorithm~\ref{Alg:solver} could be 1, if it picks $v_i$ and $v_j$ at the first timeslot. Then $T_G=1=0.5T_{\max}$. However, Lemma~\ref{lemma:1gap} also gives us that, if the initial energy of all four sensor nodes are $E\gg 1$, the lifetime gap, $T_{\max}(E)-T_G(E)$, is always 1. In this case $T_{\max}(E)=2E$ and $T_G(E)=2E-1$. Thus, the gap is negligible when $E$ is large enough, as suggested by Theorem~\ref{prop:multiply}. Therefore, when the nodal energy consumption in a timeslot is much smaller compared to the nodal battery capacity, which is generally true for sensor nodes for long term monitoring applications, the lifetime gap is negligible.
\end{remark}

Theorem~\ref{prop:multiply} shows that even though energy balancing is not equivalent to lifetime maximization in the considered network structure, the gap between the lifetime achieved by Algorithm~\ref{Alg:solver} to the maximum network lifetime is small when the initial energies of the sensor nodes are large enough compared to the energy consumption in an active timeslot. It follows that Algorithm~\ref{Alg:solver} can be used to derive good activation schedules for sensor nodes in terms of WSN lifetime.

For an illustration of the performance of Algorithm~\ref{Alg:solver}, numerical evaluations are given in Section~\ref{sec:sim}.



\section{Numerical Evaluations}\label{sec:sim}

In this section, we evaluate numerically the lifetime achieved by energy balancing, and we compare it to the maximum lifetime. Suppose the length of the pipeline under study is $L$, we normalize it to be 1 for simplicity. Then the normalized transmission range of sensor nodes is $r/L$. Two sink nodes are deployed at the end point of the pipeline, one at point 0 and the other at point 1. The sensor nodes are uniformly randomly deployed in the range of $(0,1)$. The initial energy of each node is randomly set according to a Gaussian distribution  $E_i\sim\mathcal{N}(50,5^2)$. In every timeslot, the energy of the active sensor nodes is reduced by 1, whereas the energy of other sensors remains the same. Once the residual energy of a sensor node is less than 1, it is considered as expired and is excluded from the available sensor node set $\mathcal{V}$. Once the sensor nodes in $\mathcal{V}$ become disconnected or their number  is less than $M_{\rm{cs}}$, the network has expired.

We first compare the network lifetime achieved by Algorithm~\ref{Alg:solver} to the optimal solution by solving Problem~(\ref{problem:ILP}) based on Branch-and-Bound method. As the number of nodes increases, the number of possible routes increases dramatically. Consequently, the size of the variables in the MDK Problem ~(\ref{problem:ILP}) also increases dramatically, and it becomes difficult to solve. Thus, we set the size of network to relatively small values, so that the MDK problem can be solved efficiently. The parameters of the network are as follows: the number of nodes, $N$, are randomly picked from 15 to 20, $M_{\rm{cs}}$ are randomly picked from 7 to 10, the normalized transmission range, $r/L$, is 0.25. We test 222 different random cases, among which there are 44 cases that the network lifetime by Algorithm~\ref{Alg:solver} is 1 timeslot less than the optimal, and 1 case that is 2 timeslots less than the optimal, as shown in Table~\ref{table:exhaustiveresult}. This supports our finding that balancing residual energy is effective to achieve the lifetime close to the maximum in the considered network.

\begin{table}
	\caption{Gap between lifetime achieved by energy balancing (Algorithm 1) and maximum lifetime (exhaustive search)}
	\centering
	\label{table:exhaustiveresult}
	\begin{tabular}{lllll}
		\hline
		\multicolumn{1}{c|}{Gaps (timeslots)} & \multicolumn{1}{c}{0}   & \multicolumn{1}{c}{1}  & \multicolumn{1}{c}{2} & \multicolumn{1}{c}{more} \\ \hline
		\multicolumn{1}{c|}{Number of cases}  & \multicolumn{1}{c}{177} & \multicolumn{1}{c}{44} & \multicolumn{1}{c}{1} & \multicolumn{1}{c}{0}    \\
		\hline
	\end{tabular}
	\vspace{-0.1cm}
\end{table}

%

\begin{figure}[t]
	\centering
		\includegraphics[width=0.45\textwidth]{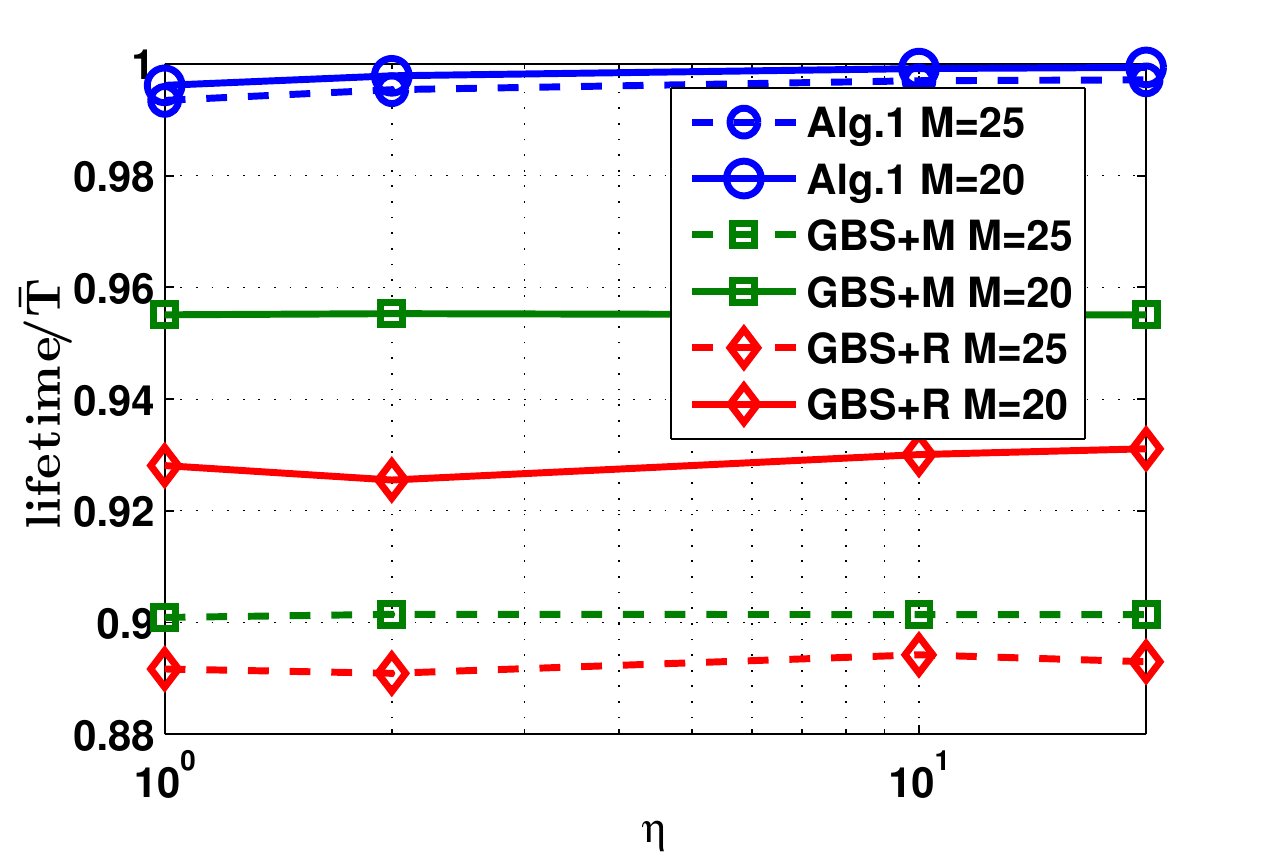}
	\caption{Comparison of the ratio of network lifetime achieved by each algorithm to the upper bound of network lifetime with  $N=100, r/L=0.15$.}
	\label{fig:multiply}  
	\vspace{-0.1cm}
\end{figure}

We further compare the performance of Algorithm~\ref{Alg:solver} to greedy based search with random (GBS+R) Algorithm and greedy based search with maximum (GBS+M)  Algorithm \cite{rong2015energy} as shown in Fig.\ref{fig:multiply}. 
We check the network lifetime when the initial energy of sensor nodes are multiplied by $\eta$. We calculate the network lifetime when $\eta$ is $1, 2, 10, 20$, respectively, and then divide the lifetime by the upper bound of the network lifetime. 
For Algorithm \ref{Alg:solver}, it is shown that the ratio of network lifetime achieved by the algorithm to the upper bound of network lifetime, $T_G(\eta)/\bar{T}(\eta)$, increases slightly as $\eta$ increases. However, such a trend does not exist for the GBS+R and GBS+M Algorithm. 

\begin{figure*}[t]
	\centering
	\subfigure[]{
		\includegraphics[width=0.32\textwidth]{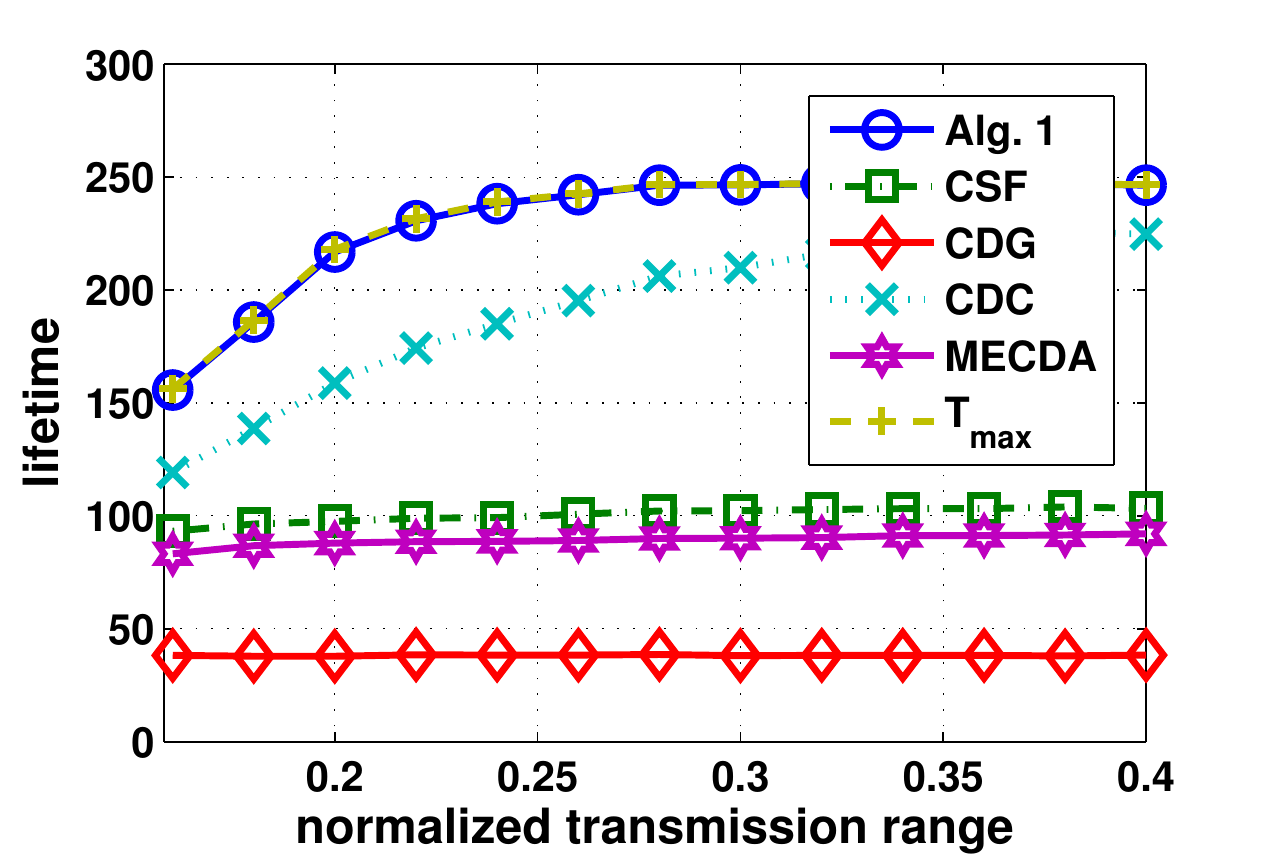}}
	\subfigure[]{
		\includegraphics[width=0.32\textwidth]{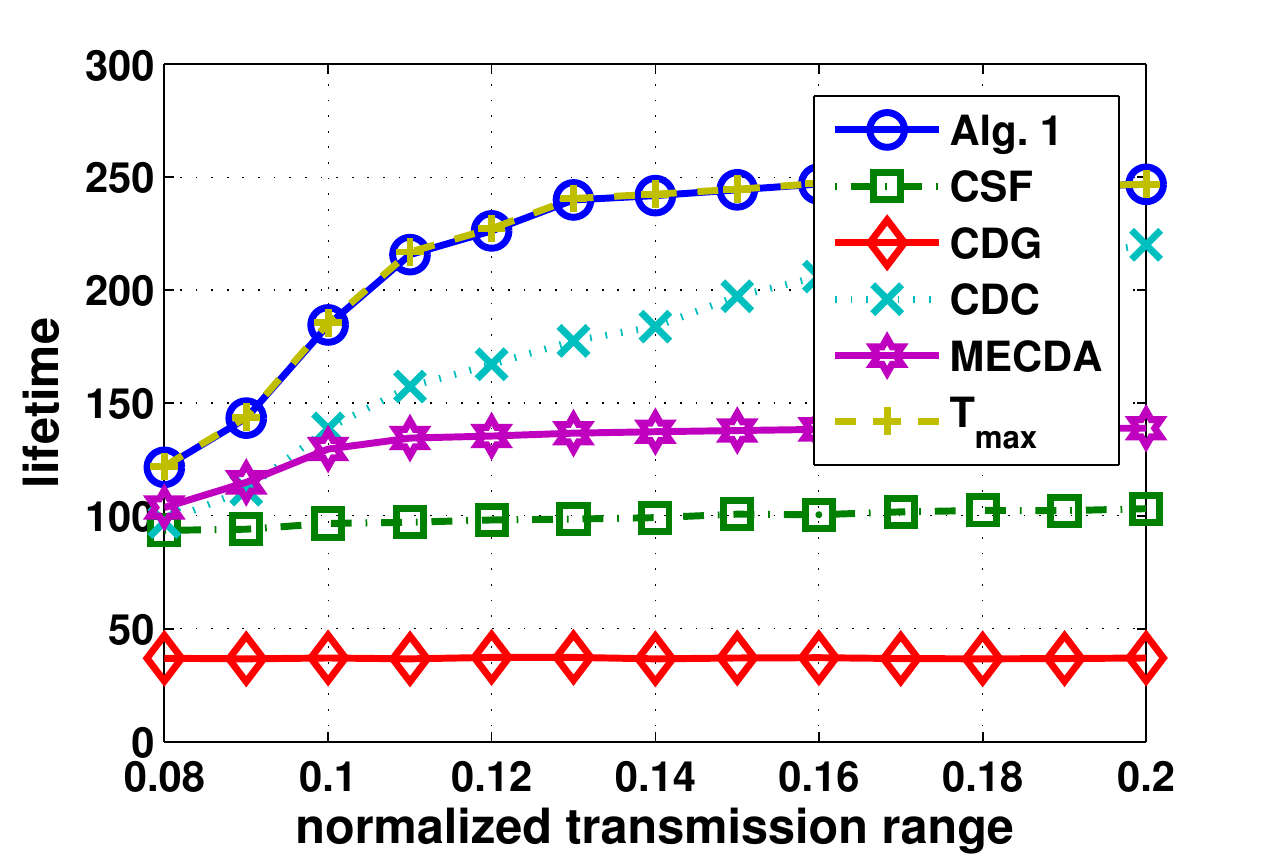}}
	\subfigure[]{
		\includegraphics[width=0.32\textwidth]{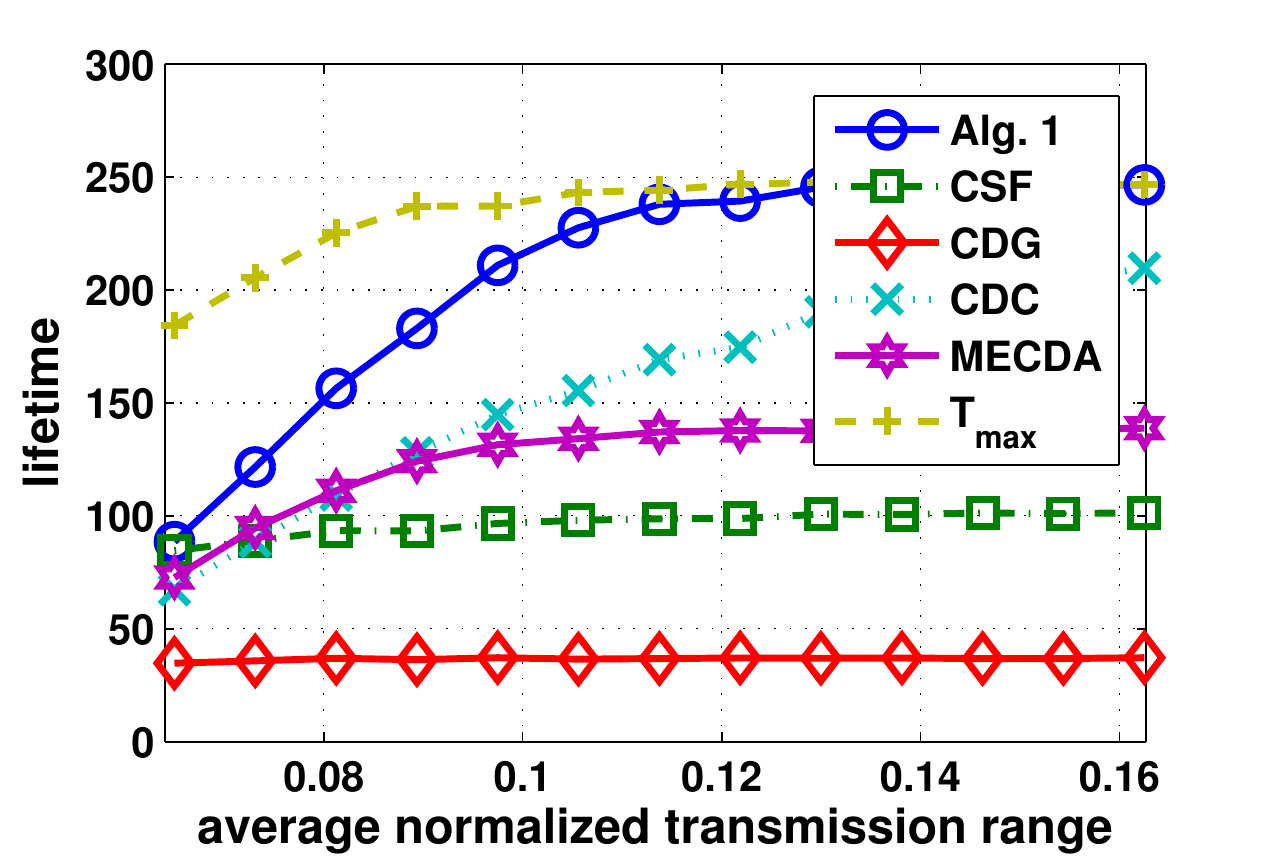}}
	\caption{Comparison of the network lifetime achieved by different algorithms with: (a) $N=50,M_{\rm{cs}}=10$, equal transmission range; (b) $N=100,M_{\rm{cs}}=20$, equal transmission range; (c) $N=100,M_{\rm{cs}}=20$, unequal transmission range}
	\label{fig:difftransmissionrange}  
	\vspace{-0.2cm}
\end{figure*}

Last, we evaluate the performance of Algorithm~\ref{SubProblemSolver} by comparing its performance to that of the state-of-art approaches, i.e., CDG \cite{luo2011compressive}, CSF \cite{xu2013efficient}, CDC \cite{liu2014CDC}, MECDA \cite{xiang2011compressed}, and also by comparing the upper bound of WSN lifetime $\bar{T}^f$ achieved according to Theorem~\ref{thm:lifetimeupperboundflow}. The results for equal transmission range are shown in Fig.~\ref{fig:difftransmissionrange} (a) and (b). The horizontal axis represents the normalized transmission range, and the vertical axis is the average WSN lifetime. The WSN lifetime achieved by Algorithm~\ref{Alg:solver} (blue solid line with circles) is very close to the upper bound (yellow dash line marked by plus) established by Theorem~\ref{thm:lifetimeupperboundflow}. It shows that the performance of Algorithm~\ref{Alg:solver} is near optimal, and the upper bound by Theorem~\ref{thm:lifetimeupperboundflow} is tight. The result of unequal transmission range is shown in Fig.~\ref{fig:difftransmissionrange} (c). In this case, the yellow dash line may not be the upper bound of the network lifetime because Assumption~\ref{assumption:transmissionrange} is not satisfied.


The results also indicate that performance achieved by Algorithm~\ref{Alg:solver} is better than that of the CDG, CSF, CDC and MECDA algorithms. Also,
the network lifetime achieved by CDG and CSF does not increase when the transmission range of sensor nodes increases. The reason is that, in these two algorithms, all sensor nodes must be constantly activated. Regarding the MECDA algorithm, since it is used to find the routing with the smallest energy consumptions, some sensor nodes are always activated until their energy is depleted. The network is then easier to become disconnected, and thus has smaller lifetime compared to the one achieved by Algorithm~\ref{Alg:solver}. Regarding the CDC algorithm, it is based on opportunistic routing. Therefore, the energy of the sensor nodes are more balanced than the case of MECDA. However, it does not guarantee minimum activation of sensor nodes in each timeslot. Therefore, it consumes more energy than Algorithm~\ref{Alg:solver} in each timeslot, and has less network lifetime. The average lifetime achieved by Algorithm~\ref{Alg:solver} is significantly longer than the CDG, CSF, CDC and MECDA approaches, which suggests the effectiveness of Algorithm~\ref{Alg:solver} in general scenarios.

Regarding the time complexity, as the complexity of  Algorithm 1 is $O(N^2)$ as discussed in our previous paper \cite{rong2015JSAC}, and recall that this algorithm determines the sensor activation for a single timeslot, the overall time complexity to achieve the approximate network lifetime based on Algorithm~\ref{Alg:solver} is $O(N^2E)$. We further test the computational time to achieve the approximate network lifetime by Algorithm~\ref{Alg:solver} and the upper bound network lifetime by testing the feasibility of Problem~\eqref{ProblemMFCC} with relaxed Constraint~\eqref{ProblemMFCC:binary}. In the settings, $N=50$ and $M_{\rm{cs}}=10$, and the nodes have the same transmission range, which is the same as in Fig.~\ref{fig:difftransmissionrange}~(a). The average computational time to retrieve the network lifetime are $1.4031$ seconds and $1.4246$ seconds for the normalized transmission range of $0.3$ and $0.4$, respectively. To calculate the lifetime upper bound $T_{\max}$, one has to test a range between the approximate lifetime $T_G$ and $\sum E_i/M_{\rm{cs}}$, which requires $4.8974$ seconds and $0.7193$ seconds for he two aforementioned cases, respectively. The computational time of calculating the lifetime upper bound reduces as the transmission range increases. The reason is that the feasible testing ranges becomes smaller and the approximate lifetime is closer to the upper bound when the node's transmission range increases.

Based on the results above, we conclude that the approach of energy balancing is effective for lifetime maximization.

\section{Conclusion} \label{sec:conclusion}

We considered a dense sensor network for monitoring a one dimensional strip area, such as a pipeline, a tunnel, or a bridge. Given that sensor node replacement is expensive and difficult, the problem of maximizing network lifetime by using compressive sensing was considered. The compressive sensing introduces a cardinality constraint, which makes the problem challenging. Thus, we characterized the performance of an approximation approach based on balancing the residual energy of sensor nodes. We proved that the resulting lifetime is at least 50\% of the optimal and that it is near optimal when the ratio of nodal initial energy to nodal energy consumptions is large enough. Simulation results showed that the ratio of the lifetime achieved by  balancing the residual energy of the nodes to the upper bound of network lifetime is close to 1 when the WSN is dense enough.

An interesting topic of future work is to study the relationship of energy balancing with lifetime maximization under cardinality constraints in a more general network structure, e.g., a WSN in 2-dimensional free spaces. Besides, deriving solution approaches that apply rounding to the results of the relaxed maximum flow problem is a promising research direction.

\appendices
\section{Proof of the equivalence of Constraints~\eqref{ProblemOPC:dynamic}-\eqref{ProblemOPC:energyconstraint} to Constraint~\ref{Problem0:Energy}}
\label{Appendix:EnergyBudget}
Constraints~\eqref{ProblemOPC:dynamic}-\eqref{ProblemOPC:energyconstraint} imply that
\begin{align*} 
0&\leq E_i(T+1)=E_i(T)-x_i(T)\\
&=E_i(T-1)-x_i(T-1)-x_i(T)\\
&=\dots\\
&=E_i(1)-\sum_{t=1}^T x_i(t)=E_i-\sum_{t=1}^T x_i(t)\,,
\end{align*}
where the first inequality comes from Constraint~\eqref{ProblemOPC:energyconstraint}, and the equalities come from Constraints~\eqref{ProblemOPC:dynamic} and~\eqref{ProblemOPC:initstate}.
Thus, Constraints~\eqref{ProblemOPC:dynamic}-\eqref{ProblemOPC:energyconstraint} can be equivalently captured by Constraint~\ref{Problem0:Energy}\,.

\section{Proof of Lemma~\ref{prop:ILP}}
\label{Appendix:MDK}
\begin{IEEEproof}
	We need to show that the solution of Problem (\ref{problem:ILP}) can be converted to the solution of Problem (\ref{ProblemOPC}), and vice versa. Suppose the solution of Problem (\ref{problem:ILP}) is $\vec{z}^*=[z^*(1),\ldots,z^*(L)]^T$. Then, we have the solution for Problem~\eqref{ProblemOPC} to be, $x_i(t)=1,\forall (t,i)\in\{(t,i)|\sum_{k=1}^{K-1}z^*(k)+1\leq t\leq \sum_{k=1}^Kz^*(k)\wedge q_K(i)=1,K=1,2,\ldots,L\}$, otherwise, $x_i(t)=0$. That is,  activate all the sensor nodes in Profile $\vec{Q}_1$ for $z^*(1)$ timeslots, and then activate all the sensor nodes in Profile $\vec{Q}_2$ for $z^*(2)$ timeslots, and so on.
	
	
	On the other hand, suppose the solution for Problem~\eqref{ProblemOPC} is $\vec{x}=\{\vec{x}(1),\ldots,\vec{x}(T)\}$. Notice that the activated sensor nodes in each timeslot must belong to a feasible activation profile, i.e., $\forall t,\exists l,Q_l=\vec{x}(t)$. Then, the solution for Problem~\eqref{problem:ILP} is $\vec{z}=[z(1),\ldots,z(L)]^T$, where $z(i)=\sum_{t=1,Q_i=\vec{x}(t)}^T 1$. This completes the proof.
\end{IEEEproof}
\begin{remark}
	Since the column in matrix $\vec{Q}$ represents a feasible activation profile, the construction of $\vec{Q}$ is equivalent to enumerating all the feasible activation profiles. It consists of two steps: 1. Search the route from $s_l$ to $s_r$; 2. Remove the routes that does not satisfy cardinality constraint. In detail, for the first step, denote $\vec{s}(v_i, k)$ the set of routes starting from $v_i$ to $s_r$ with $k$ vertices. Then, a dynamic programming based searching could proceed as $\vec{s}(v_i,k)=\bigcup_{v_j\in\mathcal{N}_-(v_i)}\vec{s}(v_j,k-1)$. Then, the set of feasible activation profiles could be represented by $\bigcup_{k\geq M+2}\vec{s}(s_l,k)$. Notice that the number of feasible activation profiles is huge for large and dense networks, and hence this approach cannot be applied in WSNs with limited storage capacity. This approach is just for performance comparison.
\end{remark}

\section{Proof of Lemma~\ref{lemma:equivalentMFCC}}
\label{Appendix:equivalentMFCC}	
	\begin{IEEEproof}
		The sketch of the proof is based on the one-to-one mapping of the constraints. Constraint~\eqref{ProblemMFCC:binary} ensues that the flows are unit flows. Together with Constraint~\eqref{ProblemMFCC:noCycle}, we have that there is at most one unit flow going out from each vertex in each timeslot. Therefore, a unit flow on edge $v_i$ to $v_j$, $u_{i,j,t}=1$, represents the activation of both nodes $v_i,v_j$ at timeslot $t$. 
		
		Constraint~\eqref{ProblemMFCC:Connectivity} represents that the output flow (the first summation) of a node should be equal to the input flow (the second summation) if the node is a sensor node. If the node is $v_0$ (the sink node $s_l$), then the difference of its output flow and its input flow should be $1$ in each timeslot. If the node is $v_{N+1}$ (the sink node $s_r$), then the difference should be $-1$ in each timeslot. Furthermore, Constraints~\eqref{ProblemMFCC:noCycle} and~\eqref{ProblemMFCC:binary} ensure that there is at most one unit flow that goes out from each vertex, i.e, there is one outgoing edge in the trail for each active node. Together with flow conservation Constraint~\eqref{ProblemMFCC:Connectivity}, there is no cyclic in the trail from $v_0$ to $v_{N+1}$. This means that the trail is a path.
		Recall Remark~\ref{rmk:connectivity} that the connectivity Constraint~\eqref{ProblemOPC:connect} is equivalent to the existence of path from $v_0$ to $v_{N+1}$. Thus, this flow conservation Constraint~\eqref{ProblemMFCC:Connectivity} with~\eqref{ProblemMFCC:noCycle} and~\eqref{ProblemMFCC:binary} is equivalent to the connectivity Constraint~\eqref{ProblemOPC:connect}. 
		
		From Constraints~\eqref{ProblemMFCC:Connectivity},~\eqref{ProblemMFCC:noCycle} and~\eqref{ProblemMFCC:binary}, the unit flow represents the activation profile in that timeslot, and $\sum_{v_j\in\mathcal{N}(v_i)} u_{i,j,t}$ is either 1 (which means $v_i$ is active) or 0 (which means $v_i$ is inactive) for each sensor node and timeslot. 
		Thus, the summation over time, $\sum_{t=1}^{T}\sum_{v_j\in\mathcal{N}(v_i)} u_{i,j,t}$, is the number of timeslots that $v_i$ is activated, and it should be smaller than $E_i$. Thus, Constraint~\eqref{ProblemMFCC:Budget} is equivalent to the energy budget Constraint~\eqref{Problem0:Energy}. Besides, the summation over nodes, $\sum_{v_i\in\mathcal{V}}\sum_{v_j\in\mathcal{N}(v_i)} u_{i,j,t}$ represents the number of nodes that is activated in a timeslot, which corresponds to the cardinality constraint. Notice that $v_l\in\mathcal{V}$, and contributes to the summation. On the other hand, even though $v_r\in\mathcal{V}$, $u_{N+1,j,t}=0$, thus, $v_r$ does not contribute to the summation. Therefore, the right hand side of Constraint~\eqref{ProblemMFCC:Cardinality} should be $M_{\rm{cs}}+1$.
		
		Given that the objectives of the two problems are identical, and the constraints are equivalent, it is concluded that Problem~\eqref{ProblemOPC} is equivalent to Problem~\eqref{ProblemMFCC}.
	\end{IEEEproof}
	
\section{The approach to solve the relaxed Problem~\eqref{ProblemMFCC}}
\label{Appendix:RelaxMFCC}

We know that given a fixed $T$, the relaxed Problem~\eqref{ProblemMFCC} is a linear optimization and thus solvable. Based on this idea, we can use a bisection approach, i.e., we can turn the problem into testing the feasibility of a linear programming problem. The idea is, given a $T$, we test whether the problem is feasible. If it is feasible, we increase $T$; otherwise we decrease it, until it converges. Notice that we have already had an upper bound of network lifetime in our previous work \cite{rong2015JSAC}, which is $\sum E_i/M_{\rm{cs}}$, the time complexity of solving the relaxed problem is  the time complexity of solving a linear programming problem, which is polynomial, multiplied by $\ln (E_i)$, and it is not high.
	

\section{Transformation to a maximum flow problem with edge capacities}
\label{Appendix:EdgeCapacity}	
\begin{figure}[t]
			\centering
			\includegraphics[width=0.2\textwidth]{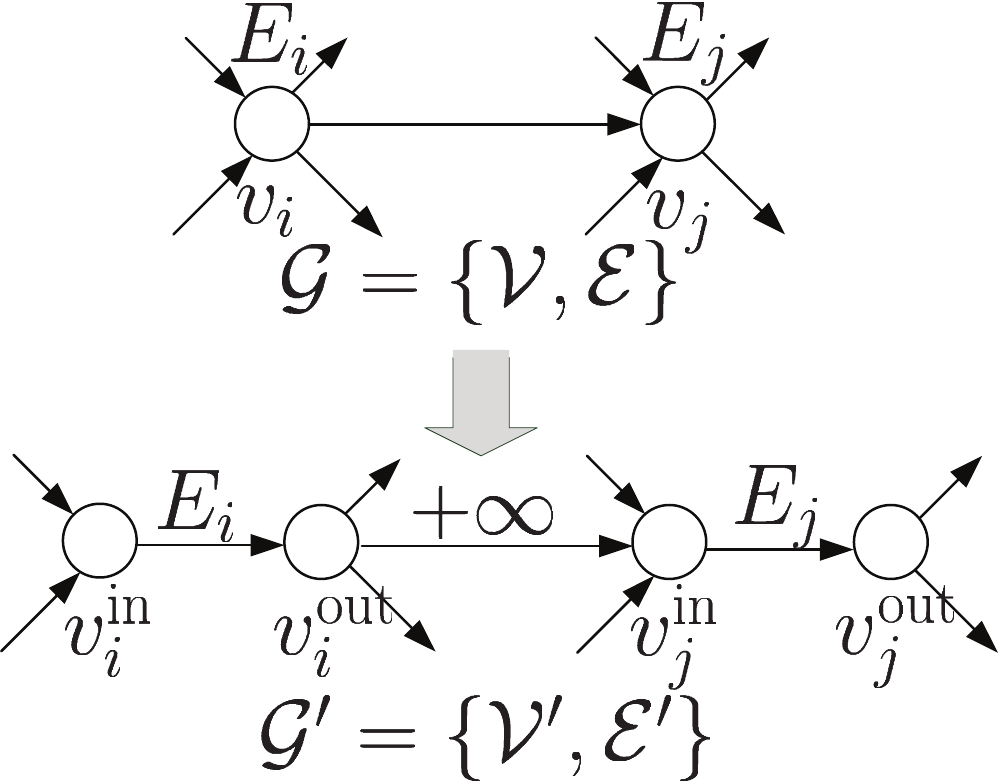}\\
			\vspace{-0.1cm}
			\caption{The transformation of a network with vertex capacity to a network with edge capacity. In the new graph $\mathcal{G}'$, the capacity $C_{ii}=E_i$, $C_{ij}=+\infty$, $C_{jj}=E_j$.}
			\label{fig:vertexcap}
			\vspace{-0.2cm}
		\end{figure}

		We need to show Problem~\eqref{ProblemMFCC} can be formulated as a maximum flow problem with edge capacities and cardinality constraints. The transformation follows the standard techniques \cite{bollobas2013modern}, as shown in Fig.~\ref{fig:vertexcap}.
		Given a network $\mathcal{G}=\{\mathcal{V},\mathcal{E}\}$, we construct a new directed graph $\mathcal{G}'=\{\mathcal{V}',\mathcal{E}'\}$. The two sink nodes in $\mathcal{V}$, $s_l$ and $s_r$ respectively, are replicated to $\mathcal{V}'$. Every sensor node $v_i$ of initial energy $E_i$ is represented by two nodes $v_i^{\text{in}}$ and $v_i^{\text{out}}$ connected by a directed arc $\langle v_i^{\text{in}},v_i^{\text{out}}\rangle$ of capacity $E_i$ in $\mathcal{G}'$. For each edge $\langle v_i,v_j\rangle \in \mathcal{E}$, if $i<j$, we construct a directed arc $\langle v_i^{\text{out}},v_j^{\text{in}}\rangle$ in $\mathcal{G}'$ with capacity $+\infty$, otherwise we construct a directed arc $\langle v_j^{\text{out}},v_i^{\text{in}}\rangle$ with capacity $+\infty$ in $\mathcal{G}'$. Thus, the vertex capacity constraints in Problem~\eqref{ProblemMFCC} turn to the edge capacity constraints. For the cardinality constraint, only the edges $\langle v_i^{\rm{in}},v_i^{\rm{out}}\rangle$ are taken into accounts. Then, the new maximum flow with edge capacity under cardinality constraint is equivalent to the Problem~\eqref{ProblemMFCC}. This completes the proof.

\section{Proof of Lemma~\ref{lemma:suboptimal}}
\label{Appendix:suboptimal}
\begin{IEEEproof}Suppose a route $\vec{R}_a$ from $s_l$ to $s_r$ contains no backward arcs, and can be divided into three parts $\vec{R}_a^1=\langle s_l,\ldots,v'_1\rangle$, $\vec{R}_a^2=\langle v'_1,\ldots,v'_2\rangle$, $\vec{R}_a^3=\langle v'_2,\ldots,s_r\rangle$. The sensor nodes in $\vec{R}_a$ are activated at $t_1$. The lifetime of an activation is $t_2-1$, i.e., at time $t_2>t_1$, we cannot find any unblocked routes from $s_l$ to $s_r$ that contains no backward arcs.  However, if we can find a route $\vec{R}_b$ that contains at least one backward arc, $\vec{R}_b$ can also be divided into three parts $\vec{R}_b^1=\langle s_l,\ldots,v'_2\rangle$, $\vec{R}_b^2=\langle v'_2,\ldots,v'_1\rangle$ and $\vec{R}_b^3=\langle v'_1,\ldots,s_r\rangle$, where $\vec{R}_b^1$ and $\vec{R}_b^3$ contain no backward arcs. If both $\langle \vec{R}_a^1,\vec{R}_b^3\rangle$ and $\langle \vec{R}_b^1,\vec{R}_a^3\rangle$ satisfy the cardinality constraint, we can pick route $\langle \vec{R}_a^1, \vec{R}_b^3\rangle$ at $t_1$ and $\langle \vec{R}_b^1,\vec{R}_a^3\rangle$ at $t_2$, such that the WSN lifetime increases from $t_2-1$ to $t_2$, which completes the proof.\end{IEEEproof}
	
\section{Proof of Lemma~\ref{lemma:backwardarc}}
\label{Appendix:backwardarc}
\begin{IEEEproof} 
	The proof is by contradiction. Suppose that a backward arc $\langle v_i^{\text{out}},v_i^{\text{in}}\rangle$ exists in $\vec{R}_b$, we divide $\vec{R}_b$ into three parts, $\vec{R}_b^1$, $\vec{R}_b^2$, $\vec{R}_b^3$, where $\vec{R}_b^1{=}\langle s_l,\ldots,v_u^{\text{out}},v_j^{\text{in}}\rangle$, $\vec{R}_b^2{=}\langle v_j^{\text{in}},\ldots, v_i^{\text{out}},v_i^{\text{in}},\ldots,v_k^{\text{out}}\rangle$, $\vec{R}_b^3{=}\langle v_k^{\text{out}},v_w^{\text{in}},\ldots,s_r\rangle$, such that backward arcs only exist in $\vec{R}_b^2$. As $\vec{R}_b$ is unblocked, $\vec{R}_b^2$ is unblocked. If $\vec{R}_b$ leads to suboptimal network lifetime, there is a route $\vec{R}_a=\langle s_l,\ldots,v_k^{\text{in}},v_k^{\text{out}},\ldots,v_i^{\text{in}},v_i^{\text{out}},\ldots,v_j^{\text{in}},v_j^{\text{out}},\ldots,s_r\rangle$ was chosen to be activated at a time $t_2\leq t_1$, where $\langle v_k^{\text{out}},\ldots,v_i^{\text{in}},v_i^{\text{out}},\ldots,v_j^{\text{in}}\rangle$ is the inverse sequence of $\vec{R}_b^2$. Similarly, we divide $R_a$  into three parts, $\vec{R}_a^1=\langle s_l,\ldots,v_{k}^{\text{out}}\rangle$, $\vec{R}_a^2=\langle v_k^{\text{out}},\ldots, v_j^{\text{in}}\rangle$, and $\vec{R}_a^3=\langle v_j^{\text{out}},\ldots,s_r\rangle$, where $\vec{R}_a^2$.
	Then we have the maximum flow increment of $\vec{R}_a^1$ and $\vec{R}_a^3$ at $t_2$ satisfies $\delta_a^1(t_2)\geq 1$, $\delta_a^3(t_2)\geq 1$, and the maximum flow increment of $\vec{R}_b^1$ and $\vec{R}_b^3$ at $t_2$ satisfies $\delta_b^1(t_2)\geq \delta_b^1(t_1+1)\geq 1$, $\delta_b^3(t_2)\geq \delta_b^3(t_1+1)\geq 1$. (Otherwise, $\vec{R}_a$ is blocked at $t_2$ and $\vec{R}_b$ is blocked at $t_1+1$.)
	
	Notice that route $\langle \vec{R}_a^1,\vec{R}_b^3\rangle$ and route $\langle \vec{R}_b^1,\vec{R}_a^3\rangle$ are not blocked at $t_2$, and they contain no backward arcs. Then, according to Line 4 to Line 5 in Algorithm \ref{SubProblemSolver}, which minimize the number of nodes to be activated in each time slot, we have that $|\vec{R}_a^1|$, $|\vec{R}_a^3|$, $|\vec{R}_b^1|$ and $|\vec{R}_b^3|$, the number of sensor nodes in $\vec{R}_a^1$, $\vec{R}_a^3$, $\vec{R}_b^1$ and $\vec{R}_b^3$ respectively, should satisfy $|\vec{R}_b^3|>|\vec{R}_a^3|$, $|\vec{R}_b^1|>|\vec{R}_a^1|$. Thus, the sensor node $v_w$ lies between $v_i$ and $v_j$, and the node $v_u$  lies between $v_k$ and $v_i$. From Assumption 1, we have $v_u\in\mathcal{N}_-(v_k)$. Further, as arc $\langle v_k^{\text{out}},v_w^{\text{in}}\rangle$ is in $\vec{R}_b^3$ $v_w\in\mathcal{N}_-(v_k)$, and hence $v_u$ and $v_w$ is connected. 
	
	Moreover, as $\vec{R}_b$ leads to the suboptimality of the network lifetime, we have that route $\langle \vec{R}_b^1,\vec{R}_a^3\rangle$ and route $\langle \vec{R}_a^1,\vec{R}_b^3\rangle$ satisfy cardinality constraint, i.e., $|\vec{R}_b^1|+|\vec{R}_a^3|\geq M_{\rm{cs}}$ and $|\vec{R}_a^1|+|\vec{R}_b^3|\geq M_{\rm{cs}}$. Together with $|\vec{R}_b^1|>|\vec{R}_a^1|$, $|\vec{R}_b^3|>|\vec{R}_a^3|$, we have that $|\vec{R}_b^1|+|\vec{R}_b^3|\geq M_{\rm{cs}}$. 
	
	Then we have an unblocked route that satisfies cardinality constraint and has no backward arcs, $\langle \vec{R}_b^{1'}, \vec{R}_b^{3'}\rangle$, at $t_1+1$ where $\vec{R}_b^{1'}$ is the route $\vec{R}_b^{1}$ without $v_j^{\text{in}}$, and $\vec{R}_b^{3'}$ is the route $\vec{R}_b^3$ without $v_k^{\text{out}}$. This comes in contradiction with that no unblocked route with only forward arcs can be found at $t_1+1$. Thus, $\vec{R}_b$ does not contain any backward arc $\langle v_i^{\text{out}}, v_i^{\text{in}}\rangle$ for all $i$, which completes the proof.
\end{IEEEproof}

\section{Proof of Lemma~\ref{lemma:1gap}}
	\label{Appendix:1gap}
	We first need a lemma that is used in the proof for Lemma~\ref{lemma:1gap}:
	\begin{lemma}\label{lemma:integers}
		Consider two positive integers $E_i$ and $E_j$ that satisfy $E_i<E_j$. For any positive integers $x$ and $y$, if $x$ and $y$ satisfy $(y-1)/E_j<x/E_i\leq y/E_j$,
		then we have $(x-1)/E_i<(y-1)/E_j$.
		\begin{IEEEproof}
			It suffices to show that $(x-1)/E_i<(y-1)/E_j$. 
			Since $x/E_i\leq y/E_j$, we have that 
			\begin{align*}
			\frac{x-1}{E_i}&<\frac{x}{E_i}-\frac{1}{E_j}\leq \frac{y}{E_j}-\frac{1}{E_j}=\frac{y-1}{E_j}\,,
			\end{align*}
			where the first inequality comes from $E_i<E_j$.
			This concludes the proof.
		\end{IEEEproof}
	\end{lemma}
	\label{appendixproof}
	Then, Lemma \ref{lemma:1gap} is proved as follows:
	
	\textit{Proof of Lemma~\ref{lemma:1gap}:}
	Suppose that backward arc $\langle v_j^{\rm{in}},v_{i}^{\rm{out}}\rangle$ with $j>i$ exists in an unblocked route $\vec{R}_b$ when the network expires at $t=t_1$. Then there exists a node $v_{h}^{\rm{out}}$ and $v_{k}^{\rm{in}}$ in $\vec{R}_b$ such that route $\langle v_{h}^{\rm{out}}, v_j^{\rm{in}}, v_{i}^{\rm{out}}, v_{k}^{\rm{in}}\rangle$ is in $\vec{R}_b$ and $h<i<j<k$. According to Assumption \ref{assumption:transmissionrange} and \ref{assumption:line}, forward arcs $\langle v_h^{\rm{out}}, v_i^{\rm{in}}\rangle$ and $\langle v_j^{\rm{out}},v_k^{\rm{in}}\rangle$ exist in $\mathcal{G}'$. As $\vec{R}_b$ leads to the suboptimality in network lifetime, we have that $v_i\in\mathcal{N}_-(v_x)$ for any node $v_x$ that $v_h\in\mathcal{N}_-(v_x)$, that $v_y\in\mathcal{N}_-(v_j)$ for any node $v_y\in\mathcal{N}_-(v_k)$, and that there is no direct edge between $v_h^{\rm{out}}$ and $v_k^{\rm{in}}$, as shown in Fig.~\ref{fig:lemma}.
	
	Then we focus on the part in $\vec{R}_b$ that contains backward arcs. Similar to the proof for Lemma \ref{lemma:backwardarc}, we again divide $\vec{R}_b$ into three parts, $\vec{R}_b^1=\langle s_l,\ldots,v_{h}^{\rm{in}},v_h^{\rm{out}}\rangle$, $\vec{R}_b^2=\langle v_h^{\rm{out}},v_j^{\rm{in}},v_{i}^{\rm{out}},v_k^{\rm{in}}\rangle$, and $\vec{R}_b^3=\langle v_k^{\rm{in}},v_k^{\rm{out}},\ldots,s_r\rangle$. If $\vec{R}_b$ causes the suboptimality in network lifetime, we have that the maximum flow increment of $\vec{R}_b^1$, $\vec{R}_b^2$, $\vec{R}_b^3$, should satisfy $\delta_b^1(t_1)\geq 1$, $\delta_b^2(t_1)\geq 1$, and $\delta_b^3(t_1)\geq 1$. Let $E_i(t)$ be the residual energy of sensor node $v_i$ before the activation at $t$-th slot, $E_i(0)=E_i$ be the initial energy of sensor node $v_i$. As $\delta_b^2(t_1)\geq 1$, we have that a route that contains $v_i$ and $v_j$ were chosen for activation at $t_2<t_1$, which means that $E_i(t_2)/E_i+E_j(t_2)/E_j\geq E_h(t_2)/E_h+E_j(t_2)/E_j$ and $E_i(t_2)/E_i+E_j(t_2)/E_j\geq E_i(t_2)/E_i+E_k(t_2)/E_k$ at $t_2$ according to Algorithm~\ref{SubProblemSolver}, which chooses the nodes having the maximum sum of normalized residual energy in every timeslot.
	
	
	Then, we divide the analysis into four cases: 1) $E_h<E_i$ and $E_k<E_j$; 2) $E_h< E_i$ and $E_k\geq E_j$; 3) $E_h\geq E_i$ and $E_k< E_j$; 4) $E_h\geq E_i$ and $E_k\geq E_j$. We will show that for case 1), there will be no flow increment; for case 2), 3) and 4), there will be at most 1.
	
	In case 1), $E_h<E_i$ and $E_k<E_j$. In the initial time when none of these four nodes have been activated, then $E_h(0)/E_h=E_i(0)/E_i=E_j(0)/E_j=E_k(0)/E_k=1$. According to Lemma \ref{lemma:integers}, we have $0/E_h<1/E_i<1/E_h$. It means that sensor node $v_h$ expires earlier than sensor node $v_i$, as the sum of residual energy of $\langle v_h,v_j\rangle$, which is $1/E_h+E_j(t)/E_j$, is always larger than that of $\langle v_i,v_j\rangle$, which is $1/E_i+E_j(t)/E_j$. Similarly, we have $0/E_k<1/E_j<1/E_k$ and hence sensor node $v_k$ expires earlier than sensor node $v_j$. Once either $v_h$ or $v_k$ expires, the route $R_b$ is blocked even when the network has not expired, which contradicts that $R_b$ is unblocked when the network expires. Thus, there will be no flow increment in this case.
	
	
	
	In case 2), $E_h<E_i$ and $E_k\geq E_j$. If $E_h\leq E_j$, with similar reason to case 1), we have that $v_h$ expires first, and then $R_b$ is blocked, which is in contradiction with that  $R_b$ is unblocked when the network expires. It means that $E_h> E_j$ as we can find an unblocked route $R_b$ when the network expires. Thus, $v_j$ expires earlier than $v_h$. If $E_j\leq E_k\leq E_i$, we have that after $v_j$ expires, $\langle v_h,v_j\rangle$ and $\langle v_i,v_j\rangle$ is blocked. The algorithm will pick $\langle v_i,v_k\rangle $ until $v_k$ expires. Then $R_b$ is blocked as $\langle v_k^{\rm{in}},v_k^{\rm{out}}\rangle$ is blocked. Consequently, $E_k>E_i>E_h> E_j$.
	Then, in the initial time when none of these four sensor nodes have been activated, the normalized residual energy of these four nodes is 1. Consequently, once $\langle v_i,v_j\rangle$ is chosen, suppose at $t_2$ when all these four nodes are not activated before, we have
	\begin{align*}
	\frac{E_i(t_2)}{E_i(0)}=\frac{E_j(t_2)}{E_j(0)}=\frac{E_k(t_2)}{E_k(0)}=\frac{E_h(t_2)}{E_h(0)}=1\,,
	\end{align*}
	\begin{align*}
	\frac{E_k(t_2+1)}{E_k(0)}>\frac{E_j(t_2+1)}{E_j(0)},\qquad
	\frac{E_h(t_2+1)}{E_h(0)}>\frac{E_i(t_2+1)}{E_i(0)}\,,
	\end{align*}
	It directly gives us that, in the next timeslot $t_2+1$,
	\begin{align}
	\frac{E_k(t_2+1)}{E_k}+\frac{E_i(t_2+1)}{E_i}>\frac{E_j(t_2+1)}{E_j}+\frac{E_i(t_2+1)}{E_i}\label{energydynamic:0}
	\end{align}
	Also, we have that
	\begin{align}
	&\frac{E_k(t_2+1)}{E_k(0)}+\frac{E_i(t_2+1)}{E_i(0)}\nonumber\\
	&\quad =\frac{E_k(t_2)}{E_k(0)}+\frac{E_i(t_2)-1}{E_i(0)}\label{energydynamic:1}\\
	&\quad=2-\frac{1}{E_i(0)}>2-\frac{1}{E_j(0)}\label{energydynamic:2}\\
	&\quad =\frac{E_h(t_2)}{E_h(0)}{+}\frac{E_j(t_2){-}1}{E_j(0)}{=}\frac{E_h(t_2{+}1)}{E_h(0)}{+}\frac{E_j(t_2{+}1)}{E_j(0)}\,,\label{energydynamic:3}
	\end{align}
	where \eqref{energydynamic:1} holds since $v_k$ is not activated but $v_i$ is activated at $t_2$, \eqref{energydynamic:2} holds since $E_i>E_j>0$, and \eqref{energydynamic:3} holds since $v_h$ is not activated but $v_j$ is activated at $t_2$.
	
	Then the algorithm will choose $\langle v_i,v_k\rangle$ instead of $\langle v_i,v_j\rangle$ and $\langle v_h,v_j\rangle$, due to \eqref{energydynamic:0} and \eqref{energydynamic:3}. After this activation, as $E_k>E_j$, we have
	$(E_i(t_2+1)-2)/E_i+(E_k(t_2+1)-1)/E_k>(E_i(t_2+1)-2)/E_i+(E_j(t_2+1)-1)/E_j$.
	It means that the normalized residual energy of route $\langle v_i,v_j\rangle$ is still smaller than that of $\langle v_i,v_k\rangle$ and $\langle v_h,v_j\rangle$. The algorithm will then pick $\langle v_i,v_k\rangle $ until the normalized residual energy of $v_k$ is smaller than that of $v_j$. When this happens, we have that the normalized residual energy of $v_i$ is smaller than that of $v_k$, and hence smaller than $v_j$ and $v_h$. It means that the algorithm will then pick $\langle v_h,v_j\rangle$. After that, the normalized residual energy of $\langle v_i,v_k\rangle$ is again larger than that of $\langle v_i,v_j\rangle$ according to Lemma \ref{lemma:integers}. This indicates that the algorithm will always pick $\langle v_i,v_k\rangle$ or $\langle v_h,v_j\rangle$ instead of $\langle v_i,v_j\rangle$. Consequently,  the maximum flow increment of $\langle v_j^{\rm{in}}, v_{i}^{\rm{out}}\rangle$ in this case is 1 as $\langle v_i,v_j\rangle$ was chosen only once.
	
	
	As case 3) is symmetric to case 2), we have the maximum flow increment of $\langle v_j^{\rm{in}},v_{i}^{\rm{out}}\rangle$ in this case is also 1.
	
	In case 4), $E_h\geq E_i$ and $E_k\geq E_j$. We have for any positive integer $x$, $x/E_i\geq x/E_h$, and $x/E_j\geq x/E_k$. In the initial time when none of these four sensor nodes have been activated, the normalized residual energy of these four nodes is equal to 1. Hence, once $\langle v_i,v_j\rangle$ is chosen, suppose at $t_2$, we have
	\begin{align}
	&\frac{E_h(t_2+1)}{E_h(0)}{>}\frac{E_i(t_2+1)}{E_i(0)},\label{batterystate:5}\\
	&\frac{E_k(t_2+1)}{E_k(0)}{>}\frac{E_j(t_2+1)}{E_j(0)}\,,
	\end{align}
	and thus
	\begin{align*}
	&\frac{E_i(t_2+1)}{E_i(0)}+\frac{E_j(t_2+1)}{E_j(t_2+1)}<\frac{E_i(t_2+1)}{E_i(0)}+\frac{E_k(t_2+1)}{E_k(0)}\,,\\
	&\frac{E_i(t_2+1)}{E_i(0)}+\frac{E_j(t_2+1)}{E_j(t_2+1)}<\frac{E_h(t_2+1)}{E_h(0)}+\frac{E_j(t_2+1)}{E_j(0)}\,.
	\end{align*}
	Thus, the algorithm will choose $\langle v_i,v_k\rangle $ or $\langle v_h,v_j\rangle$ instead of $\langle v_i,v_j\rangle$. Due to the symmetry, the analysis when it picks $\langle v_i,v_k\rangle$ is similar to that if it picks $\langle v_h,v_j\rangle$. Thus, we only analyze the case if it picks $\langle v_i,v_k\rangle$.
	
	Suppose that the algorithm  picks $\langle v_i,v_k\rangle$, then the normalized residual energy of sensor node $v_i$ and $v_k$ becomes $(E_i(t_2+1)-1)/E_i$ and $(E_k(t_2+1)-1)/E_k$.
	Suppose $(E_k(t_2+1)-1)/E_k>E_j(t_2+1)/E_j$, we have that the normalized residual energy of route $\langle v_i,v_j\rangle$ is still less than that of $\langle v_i,v_k\rangle$ and $\langle v_h,v_j\rangle$. Otherwise, we have 
	\begin{align*}
	&\frac{E_j(t_2+1)-1}{E_j}+\frac{E_h(t_2+1)}{E_h}\\
	&\quad > \frac{E_j(t_2+1)-1}{E_j}+\frac{E_i(t_2+1)}{E_i}\\
	&\quad > \frac{E_j(t_2+1)-1}{E_j}+\frac{E_i(t_2+1)-1}{E_i}\,,
	\end{align*}
	where the first inequality comes from~\eqref{batterystate:5}. The summation of residual energy of $v_j$ and $v_h$ is larger than that of $v_i$ and $v_j$.
	Thus, the algorithm will pick route $\langle v_h,v_j\rangle$. Then the normalized residual energy of these four sensor nodes are $(E_h(t_2+1)-1)/E_h$, $(E_i(t_2+1)-1)/E_i$, $(E_j(t_2+1)-1)/E_j$ and $(E_k(t_2+1)-1)/E_k$. As $1/E_i>1/E_h$, $1/E_j>1/E_k$, we have
	\begin{equation}\label{energydynamic:4}
	\begin{split}
	&\frac{E_i(t_2+1)-1}{E_i(0)}+\frac{E_j(t_2+1)-1}{E_j(0)}\\
	&\quad=\frac{E_i(t_2+1)}{E_i(0)}-\frac{1}{E_i(0)}+\frac{E_j(t_2+1)-1}{E_j(0)}\\
	&\quad<\frac{E_h(t_2+1)}{E_i(0)}-\frac{1}{E_h(0)}+\frac{E_j(t_2+1)-1}{E_j(0)}\\
	&\quad=\frac{E_h(t_2+1)-1}{E_h(0)}+\frac{E_j(t_2+1)-1}{E_j}\,,
	\end{split}
	\end{equation}
	and similarly
	\begin{equation}\label{energydynamic:5}
	\begin{split}
	&\frac{E_i(t_2+1)-1}{E_i(0)}+\frac{E_j(t_2+1)-1}{E_j(0)}\\
	&\quad<\frac{E_i(t_2+1)-1}{E_i(0)}+\frac{E_k(t_2+1)-1}{E_k(0)}\,.
	\end{split}
	\end{equation}
	\eqref{energydynamic:4} and \eqref{energydynamic:5} indicate that the algorithm will always pick $\langle v_i,v_k\rangle$ or $\langle v_h,v_j\rangle$ instead of $\langle v_i,v_j\rangle$, until $v_i$ or $v_j$ expires. Consequently, the maximum flow increment of $\langle v_j^{\rm{in}}, v_{i}^{\rm{out}}\rangle$ in this case is 1, as $\langle v_i,v_j\rangle$ was only picked once.
	
	
	To sum up, the maximum flow increment of $\langle v_j^{\rm{in}},v_{i}^{\rm{out}}\rangle$ is 1 for Cases 2) to 4), and it is 0 for Case 1). Thus, the maximum flow increment of  $\langle v_j^{\rm{in}},v_{i}^{\rm{out}}\rangle$ is 1.
	\endproof

\bibliographystyle{IEEEtran}
\bibliography{sensing2}

\end{document}